\documentclass{IOS-Book-Article}

\usepackage{graphicx}
\usepackage{multicol}
\usepackage{amsmath}
\usepackage{array}
\usepackage{mdwmath}
\usepackage{mdwtab}
\usepackage{eqparbox}
\usepackage{tabularx}
\usepackage[font=footnotesize]{subfig}

\usepackage{stfloats}
\usepackage{url}
\usepackage{booktabs}
\usepackage{hyperref}    
\usepackage{algorithm}
\usepackage{algorithmicx}
\usepackage{varwidth}
\usepackage{algpseudocode}
\usepackage{amssymb}
\usepackage{mathrsfs}
\usepackage{float}
\usepackage{multirow}
\usepackage{hhline}

\usepackage{caption}
\def\hb{\hbox to 10.7 cm{}}

\begin{document}

\pagestyle{headings}
\def\thepage{}

\begin{frontmatter}
\title{High-Performance Massive Subgraph Counting using Pipelined Adaptive-Group Communication}
\markboth{}{May 2018\hb}

\runningtitle{}
\author[A]{\fnms{Langshi} \snm{Chen}},
\author[A]{\fnms{Bo} \snm{Peng}},
\author[A]{\fnms{Sabra} \snm{Ossen}},
\author[B]{\fnms{Anil} \snm{Vullikanti}},
\author[B]{\fnms{Madhav} \snm{Marathe}},
\author[A]{\fnms{Lei} \snm{Jiang}},
\author[A]{\fnms{Judy} \snm{Qiu}}
\runningauthor{}
\address[A]{Dept. of Intelligent Systems Engineering, Indiana University\\\{lc37,bpeng,jiang60,xqiu\}@indiana.edu, sossen@iu.edu}
\address[B]{Biocomplexity Institute and Dept. of Computer Science, Virginia Tech \\ \{vsakumar,mmarathe\}@vt.edu}

\begin{abstract}
Subgraph counting aims to count the number of occurrences of a subgraph  $T$ (aka as a template) in a given graph $G$. The basic problem has found applications in diverse domains. The problem is known to be computationally challenging -- the complexity grows both as a function of $T$ and $G$. Recent applications have motivated solving such problems on massive networks with billions of vertices. 

In this chapter, we study the subgraph counting problem from a parallel computing perspective. We discuss efficient parallel algorithms for approximately resolving subgraph counting problems by using the color-coding technique. We then present several system-level strategies to substantially improve the overall performance of the algorithm in massive subgraph counting problems. We propose: 1) a novel pipelined Adaptive-Group communication pattern to improve inter-node scalability, 2) a fine-grained pipeline design to effectively reduce the memory space of intermediate results, 3) partitioning neighbor lists of subgraph vertices to achieve better thread concurrency and workload balance.  
Experimentation on an Intel Xeon E5 cluster shows that our implementation
achieves 5x speedup of performance compared to the state-of-the-art work while reduces the peak memory utilization by a factor of 2 on large templates of 12 to 15 vertices and input graphs of 2 to 5 billions of edges.

\end{abstract}

\end{frontmatter}
\markboth{May 2018\hb}{May 2018\hb}

\section{Introduction}
\label{sec:intro}
Subgraph analysis in massive graphs is a fundamental task that arises in numerous applications, including social network analysis 
\cite{chen:icdm16}, uncovering network motifs (repetitive subgraphs) in gene regulatory networks in bioinformatics \cite{milo2002network}, indexing graph databases \cite{khan:sigmod11}, optimizing task scheduling in infrastructure monitoring, and detecting events in cybersecurity. Many emerging applications often require one to solve the subgraph counting problem for very large instances. 

Given two graphs---a subgraph $T$ on $k$ vertices (also referred to as a template),  and a graph $G$ on $n$ vertices and $m$ edges as input, some of the commonly studied questions related to the subgraph analysis include: 1) Subgraph-existence: determining whether $G=(V,E)$ contains a subgraph that 
is isomorphic to template $T$, 2) Subgraph-counting: counting the number of such subgraphs, 
3) Frequent-subgraphs: finding subgraphs that occur frequently in $G$, and 4) Graphlet frequency: computing the frequency 
Distribution (GFD) of a set of templates ${\cal T}$, 
i.e. for each template $T \in {\cal T}$ count the number of occurrences of $T$ in in $G$. Some of the commonly studied templates are paths and trees, and 
we focus on the detection and counting versions of the non-induced subgraph isomorphism problem for paths and trees (these problems  are formally defined in Section \ref{sec:prelim}). 
Tree template counting can also be used as a kernel to estimate the GFD in a graph. For instance, Bressan et al.~\cite{bressan_counting_2017} show that a well-implemented tree template counting kernel can push the limit of the state-of-the-art GFD in terms of input graph size and template size. These problems are recognized to be NP-hard even for very simple templates, and the best algorithms for computing exact counts of a $k$-vertices template from a $n$-vertices graph has a complexity of $\Omega(n^{k/2})$ \cite{vassilevska2009finding}. 

This motivates the use of approximation algorithms, and several techniques have been developed for subgraph counting problems. These have been based on the idea of \emph{fixed parameter tractable} algorithms, whose execution time is exponential in the template size $k$ 
but polynomial in the number of vertices, $n$---this is one of the standard approaches for dealing with NP-hard problems (see~\cite{flum2004parameterized,curticapean2014complexity}). Two powerful classes of techniques that have been developed for subgraph counting are: 1) color-coding~\cite{alon_color-coding_1995}, which was the first fixed parameter technique for counting paths and trees, with a running time and space $O((2e)^km)$ and $O(2^kn)$, respectively,
and 2) multilinear detection \cite{koutis:icalp08, williams2009finding}, which is based on a reduction of the subgraph detection problem to detecting multilinear terms in multivariate polynomials. This approach reduces the time and space to $O(2^km)$ and $O(k)$, respectively. However, finding the actual subgraphs requires additional work \cite{bjorklund:esa14}.

Our focus is on parallel algorithms and implementations for subgraph detection and counting. Parallel versions of both the color-coding~\cite{zhao_sahad:_2012, slota_parallel_2015} and  multilinear detection techniques have been developed~\cite{ekanayake:ipdps18}. Though the multilinear detection technique has several benefits in terms of time and space over color-coding, it is still more involved (than color-coding) when it comes to finding the subgraph embeddings. Additionally, the color-coding technique has been extended to subgraphs beyond trees, specifically, those with treewidth more than 1---such subgraphs are much more complex than trees and can contain cycles
\cite{chakaravarthy_subgraph_2016}. Therefore, efficient parallelization of the color-coding technique remains a very useful objective, and will be the focus of our work.

The current parallel algorithms for  color-coding have been either implemented with MapReduce (SAHAD~\cite{zhao_sahad:_2012}) or with MPI (FASCIA~\cite{slota_parallel_2015}). However, both methods suffer from significant communication overhead and large memory footprints, which prevents them from scaling to templates with more than 12 vertices.
%
%
We focus on the problem of counting tree templates (referred to as treelets), identify the bottlenecks of scaling, and design a new approach for parallelizing color-coding. We aim to address the following computation challenges:
\begin{itemize}
	\item \textbf{Communication:} Many graph applications are based on point-to-point communication, having the unavailability of high-level communication abstraction that is adaptive for irregular graph interactions. 
	\item \textbf{Load balance:} Sparsity of graph leads to load imbalance of computation.  
	\item \textbf{Memory:} High volume of intermediate data, due to large subgraph template (big model), causes intra-node high peak memory utilization at runtime.
\end{itemize}
We investigate computing capabilities to run subgraph counting at a very large scale, and we propose the following solutions: 
\begin{itemize}
    \item \textbf{Adaptive-Group communication} with a data-driven pipeline design to interleave computation with communication.
    \item \textbf{Partitioning neighbor list} for fine-grained task granularity to alleviate load imbalance at
    thread level within a single node.
    \item \textbf{Intermediate data partitioning} with sliced and overlapped workload in the pipeline to reduce peak memory utilization.  
\end{itemize}

We compare our results with the state-of-the-art MPI Fascia implementation \cite{slota_parallel_2015} and show applicability of the proposed method by counting large treelets (up to 15 vertices) on massive graphs (up to 5 billion edges and 0.66 billion vertices). 	

The rest of the chapter is organized as follows. Section \ref{sec:prelim} introduces the problem, color-coding algorithm and scaling challenges. 
Section~\ref{sec:approach} presents our approach on 
Adaptive-Group communication as well as a neighbor list partitioning technique at thread level. 
Section~\ref{sec:eval} contains experimental analysis of our proposed methods and the performance improvements. Related works and our conclusion could be found in Section \ref{sec:related_work} and \ref{sec:conclusion}.   

\section{Background}
\label{sec:prelim}

Let $G=(V, E)$ denote a graph on the set $V$ of nodes and set $E$ of edges.
We say that a graph $H=(V_H, E_H)$ is a \emph{non-induced subgraph} of $G$ if
$V_H\subseteq V$ and $E_H\subseteq E$. We note that there may be other edges
in $E-E_H$ among the nodes in $V_H$ in an induced embedding.
A template graph $T = (V_T, E_T)$ is said to be isomorphic to a non-induced
subgraph $H = (V_H,E_H)$ of $G$ if there exists a bijection
$f: V_T \rightarrow V_H$ such that for each edge $(u, v) \in E_T$, we have $(f(u), f(v))\in E_H$.
In this case, we also say that $H$ is a non-induced embedding of $T$.

\noindent
\textbf{Subgraph-counting problem.} \ Given a graph $G=(V, E)$ and a template $T=(V_T, E_T)$, our objective is to count the number of non-induced embeddings of $T$ in $G$, which is denoted by $\#emb(T, G)$. 

Given $\epsilon, \delta\in(0, 1)$, we say that a randomized algorithm $\mathcal{A}$ gives an $(\epsilon, \delta)$-approximation if $\Pr[|\mathcal{A}(T, G) - \#emb(T, G)| > \epsilon\#emb(T, G)] < \delta$. In this chapter, we will focus on obtaining an $(\epsilon, \delta)$-approximation to $\#emb(T, G)$.

\subsection{The Color-coding Technique}
Color-coding is a randomized approximation algorithm, which estimates the number of tree-like embeddings in $O(c^k\text{poly}(n))$ with a tree size $k$ and a constant $c$. 
We briefly describe the key ideas of the color-coding technique here, since our algorithm involves a parallelization of it.

\vspace{0.2 cm}
\noindent
\textbf{Counting colorful embeddings.}
The main idea is that if we assign a color $col(v)\in\{1,\ldots,k\}$ to each node $v \in G$,
``colorful'' embeddings, namely those in which each node has a distinct color, can be
counted easily in a bottom-up manner. 

For a tree template $T=(V_T, E_T)$, let $\rho(T)$ denote its root, which can be picked arbitrarily. Then $T(v)$ denote a template $T$ with root $v=\rho(T)$.
Let $T'$ and $T''$ denote
the subtrees by cutting edge $(\rho(T), u)$ from $T$. We pick
$\rho(T') = \rho(T)$ and $\rho(T'')=u$.
Let $C(v, T, S)$ denote the number of colorful embeddings of $T$ with vertex
$v \in V_G$ mapped to the root $\rho(T)$, and using the color set $S$, where $|V_{T}| = |S|$.
Then, we can compute $C(v, T, S)$ using dynamic programming with the following recurrence. \\
\begin{align}
C(v, T, S) & = \frac{1}{d}\sum_{u\in N(v)} \sum_{S=S_1\cup S_2} C(v, T', S_1) \cdot C(u, T'', S_2)
\end{align}
where $d$ accounts for the over-counting, as discussed in \cite{alon_color-coding_1995}.

\begin{figure}[!ht]
  \centering
  \subfloat[Template Partition]{\includegraphics[width=0.3\linewidth]{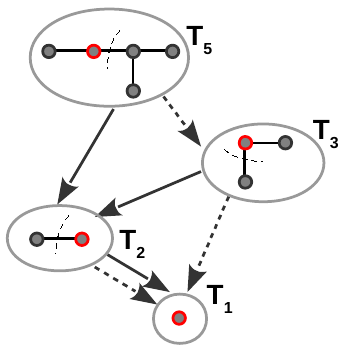}}
  \hspace{2ex}
  \subfloat[Colorful Embedding Counting]{\includegraphics[width=0.55\linewidth]{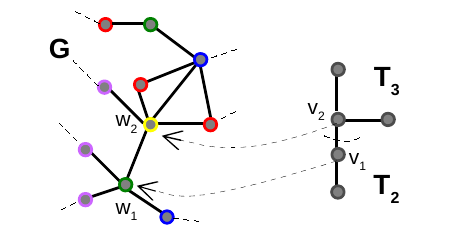}}
  \caption{An example showing the two main steps of color-coding with tempate $T_5$.} 
  \label{fig:cc_example}
\end{figure}

Figure \ref{fig:cc_example} (a) shows how the problem is decomposed into smaller sub-problems. In this partition process, an arbitrary vertex  is picked up as the root which is marked in red, then one edge of it is removed, splitting tree $T$ into two small sub-trees. The arrow lines denote these split relationships, with the solid line pointing to the sub-tree with the root vertex and dotted line to the other. This process runs recursively until the tree template has only one vertex, $T_1$. 
Figure \ref{fig:cc_example} (b) shows an example of the colorful embedding counting process which demonstrates the calculation on one neighbor of the root vertex. 
Here, tree template $T_5$ is split into sub templates $T_2$ and $T_3$, 
in order to count $C(w_1 , T_5(v_1), S)$, or the number of embeddings of $T_5(v_1)$ rooted at $w_1$, using color set $S=\{red, yellow, blue, green, purple\}$, 
we enumerate over all valid combination of sub color sets on $T_2$ and $T_3$.
For $S_1=\{g,p\}, S_2=\{y,r,b\}$, we have
$C(w_1 , T_2(v_1), \{g,p\}) = 2$ and $C(w_2 , T_3(v_2), \{y,r,b\}) = 2$, and 
for $S_1=\{g,b\}, S_2=\{y,r,p\}$, we have
$C(w_1 , T_2(v_1), \{g,b\}) = 1$ , $C(w_2 , T_3(v_2), \{y,r,p\}) = 2$. 
As $T_5$ can be constructed by combinations of these sub trees, 
$C(w_1 , T_5(v_1), S)$ equals to the summation of the multiplication of the count of the sub trees, and results in $2\times2 + 1\times2 = 6$.
In this example, the combination of two sub-trees of $T_5$ uniquely locates a colorful embedding. But for some templates, some subtrees are isomorphic to each other when the root is removed. E.g., for $T_3$ in Figure~\ref{fig:cc_example} (a), the same embedding will be over-counted for 2 times in this dynamic programming process. 

\noindent
\textbf{Random coloring.}
The second idea is that if the coloring is done randomly with $k=|V_T|$ colors, there will be a reasonable probability that an embedding is colorful, i.e., each of its nodes is marked by a distinct color. Specifically, an embedding $H$ of $T$ is colorful with probability $\frac{k!}{k^k}$.
Therefore, the expected number of colorful embeddings is $n(T, G)\frac{k!}{k^k}$. Alon et al. \cite{alon_color-coding_1995} show that this estimator has bounded variance, which can be used to efficiently estimate the number of embeddings, denoted as $n(T, G)$. Algorithm \ref{alg:sequential} describes the sequential color-coding algorithm.
\begin{algorithm}[htbp]
  \caption{Sequential color-coding algorithm.}
  \label{alg:sequential}
  \begin{algorithmic}[1]
\State \textbf{Input:} Graph $G=(V, E)$, a template $T=(V_T, E_T)$, and parameters $\epsilon$, $\delta$
\State \textbf{Output:} A $(1\pm\epsilon)$- approximation to $\#emb(T, G)$ with probability of at least $1-\delta$
\State Let $Niter=O(\frac{e^k\log(1/\delta)}{\varepsilon^2})$ and $k=|V_T|$
\For{$j=1$ to $Niter$}
\For{$v\in V$}
\State Pick a color $col(v)\in S=\{1,\ldots,k\}$ uniformly at random
\EndFor
\State Partition $T$ into subtrees recursively to form $\mathcal{T}$.
\For{$v\in V$, $T_i\in\mathcal{T}$ and subset $S_i\subseteq S$, with $|S_i|=|T_i|$}
\State Compute
\begin{equation}
\label{eq:color-coding-dp}
\begin{array}{lcl}
\hspace*{-0.05in}
C(v, T_i, S_i) = \displaystyle\frac{1}{d}\sum\limits_{u}
\displaystyle\sum C(v, T_i', S_i') \cdot C(u, T_i'', S_i''),\end{array}
\end{equation}
where $T_i$ is partitioned into trees $T'_i$ and $T''_i$ in $\mathcal{T}$, and $d$ is the over counting factor for $T_i$, as described in \cite{alon_color-coding_1995}.
\EndFor
\State 
Compute $C^{(j)}$, the number of colorful embeddings of $T$ in $G$ for the $j$th coloring as
\begin{equation}
\label{eq:color-coding-sum}
\begin{array}{lcl}
C^{(j)} = \frac{k^k}{k!}\sum_{ v \in V} C(v, T(\rho), S)
\end{array}
\end{equation}
\EndFor
\State Partition the $Niter$ estimates $C^{(1)},...,C^{(Niter)}$ into $t=O(\log(1/\delta))$ sets of equal size. Let
$Z_j$ be the average of set $j$. Output the median of $Z_1,...,Z_t$.
\end{algorithmic}
\end{algorithm}

\vspace{0.2 cm}
\noindent
\textbf{Distributed color-coding and challenges}.
As color-coding runs $Niter$ independent estimates in the outer loop at line 4 in the sequential Algorithm \ref{alg:sequential}, 
it is straightforward to implement the outer loop at line 4 in a parallel way. However,  
if a large dataset cannot fit into the 
memory of a single node, the algorithm must partition the dataset over multiple nodes and parallelize the 
inner loop at line 8 of Algorithm~\ref{alg:sequential} to exploit computation horsepower from more cluster nodes. 
Nevertheless, vertices partitioned on each local node requires count information of their neighbor $u$ located on remote cluster nodes, which brings communication 
overhead that compromises scaling efficiency. Algorithm~\ref{alg:distriColorCoding} uses a collective all-to-all operation to communicate count information 
among processes and updates the counts of local vertices at line 16. This standard communication pattern 
ignores the impact of growing template size, which exponentially increases communication cost and reduces the parallel scaling efficiency. Moreover, skewed distribution of neighbor vertices on local cluster nodes will generally cause workload imbalance among processes and produce a ``straggler" to slow down the collective communication operation. Finally, it requires each local node to hold all the transferred count information in memory 
before starting the computation stage on the remote data, resulting in a high peak memory utilization on a single cluster node and becoming a bottleneck in scaling out the distributed color-coding algorithm. 

\begin{algorithm}[htbp]
\caption{Distributed color-coding algorithm}
\label{alg:distriColorCoding}
    \begin{algorithmic}[1]
	\State \begin{varwidth}[t]{\linewidth}
		\textbf{Input}: Graph $G(V,E)$, Template $T$, \par
		 Processes $P$, Color number $k$ \par
		 $\delta$ and $\varepsilon$ are parameters that control approximation quality \par
		\textbf{Output}: A $(1\pm\varepsilon)$- approximation to $\#emb(T, G)$ with probability of at least $1-\delta$
	\end{varwidth}
	\State $G(V,E)$ is randomly partitioned into $P$ processes
	\State $T$ is partitioned into subtemplates $T_i \in \mathcal{T}$ 
	\State $\rho$ is the root of $T$
    \State $Niter=O(\frac{e^k\log(1/\delta)}{\varepsilon^2})$
    \State $S=\{1,\ldots,k\}$ is a color set
	\For{it=1 to $Niter$} \Comment{Out-loop iterations}
	\For{Each process $0\leq p \le P$} \Comment{Process-level parallelism}
	\State Color local graph $G_p(V,E)$	
	\ForAll{$T_i \in \mathcal{T}$ in reverse order of partitioning}
	\State select subset $S_i\subseteq S$, with $|S_i|=|T_i|$
	\ForAll{$v \in G_p(V,E)$} \Comment{Thread-level parallelism}
	\State Compute $C_p(v,T_i, S_i)$ from neighbor vertices of $v$ within process $p$
\EndFor
\State 
\begin{varwidth}[t]{\linewidth}
	Process $p$ All-to-All exchanges local counts\par 
	$C_p(,T_i,S_i)$ with other processes 
\end{varwidth}
\ForAll{$v \in G_p(V,E)$} \Comment{Thread-level parallelism}
\State 
\begin{varwidth}[t]{\linewidth}
	Update $C_p(v,T_i,S_i)$ by computing received \par
	neighbor vertices of $v$ from other processes
\end{varwidth}
\EndFor
\EndFor
\EndFor
\EndFor
\State Compute $C^{(j)}$, the number of colorful embeddings of $T$ in $G$ for the $j$th coloring as
$C^{(j)} = \frac{k^k}{k!} \sum_{v\in V}c(v,T(\rho),S)$
\State
Partition the $Niter$ estimates $C^{(1)},...,C^{(Niter)}$ into $t=O(\log(1/\delta))$ sets of equal size. Let
$Z_j$ be the average of set $j$. Output the median of $Z_1,...,Z_t$.
\end{algorithmic}
\end{algorithm}

\section{Scaling Distributed Color-coding}
\label{sec:approach}

To address the challenges analyzed in Section~\ref{sec:prelim}, we propose a novel node-level communication scheme named Adaptive-Group in Section~\ref{sec:adaptive_group}, and a fine-grained thread-level optimization called neighbor list partitioning in Section~\ref{sec:neighbor_list_partition}. 
Both of the approaches are implemented as a subgraph counting application to our open source project Harp-DAAL~\cite{chen_benchmarking_2017}\cite{harpdaalweb}.

\subsection{Harp-DAAL}
\label{sec:harpdaal}

Harp-DAAL is an ongoing effort of running data-intensive workloads on HPC clusters and becoming a High Performance Computing Enhanced Apache Big Data Stack (HPC-ABDS). Being extended from Apache Hadoop, Harp-DAAL provides users of MPI-like programming model besides the default MapReduce paradigm. Unlike Apache Hadoop, Harp-DAAL utilizes the main memory rather than the hard disk to store intermediate results, and it implements 
a variety of collective communication operations optimized for data-intensive machine learning workloads 
\cite{Zhang2015a}\cite{Zhang2016}\cite{peng_harplda+:_2017}. Furthermore, Harp-DAAL provides hardware-specific acceleration via an integration of Intel's Data Analytics and Acceleration Library 
(Intel DAAL) \cite{inteldaalweb} for intra-node computation workloads. Intel DAAL is an open source project, 
and we contribute the optimization codes of this work as re-usable kernels of Intel DAAL. 

\subsection{Adaptive-Group Communication}
\label{sec:adaptive_group}
\begin{figure}[!ht]
    \centering
    \includegraphics[width=\linewidth]{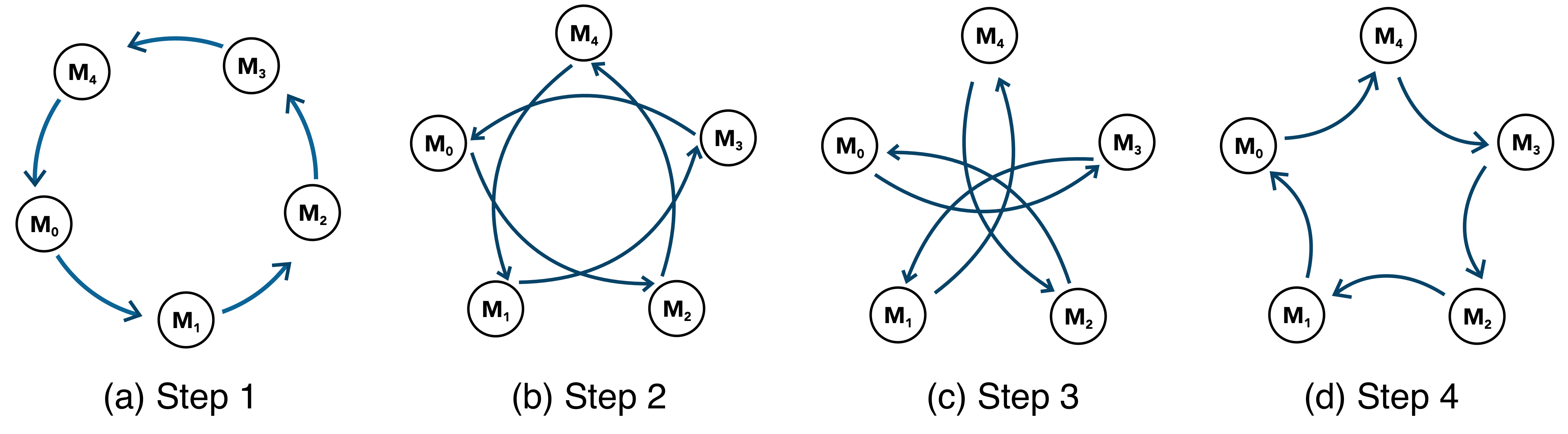}
    \caption{An example of ring-ordered steps in the Adaptive-Group communication}
    \label{fig:ring-order}
\end{figure}
Adaptive-Group is an interprocess communication scheme based on the concept of communication group. 
Given $P$ parallel computing processes, each process $p$ belongs to a communication group where it has data dependencies, i.e., sending/receiving data, with other processes in the group. In an all-to-all operation, such as MPI\_Alltoall, each process $p$ communicates data with all the other processes in a collective way, namely all processes are 
associated to a single communication group with size $P$. In Adaptive-Group communication, the collective communication is  
divided into $W$ steps, where each process $p$ only communicates with processes belonging to a communication group of size $m$ at each step $w$. The size $m$ and the number of total steps $W$ are both configurable on-the-fly and adaptive to computation overhead, load balance, and memory utilization of irregular problems like subgraph counting.\par 
A routing method is required to guarantee that no missing and redundant data transfer occurs during all the $W$ steps. 
Figure~\ref{fig:ring-order} illustrates such a routing method, where the all-to-all operation among 5 processes 
is decoupled into 4 steps, and each process only communicates with two other processes within a communication group of size 3 at each step. 
Line 3 to 14 of Algorithm~\ref{alg:adaptive-group} gives out the pseudo code of Adaptive-Group 
communication that implements the routing method in Figure~\ref{fig:ring-order}. Here the communication is adaptive to the template size $|T|$. With
a large template size $|T|$, the algorithm adopts the routing method in Figure~\ref{fig:ring-order} with a communication group size of 3, while it 
switches to the traditional all-to-all operation if the template size is small.
\begin{algorithm}[!ht]
\caption{Adaptive-Group in distributed color-coding}
\label{alg:adaptive-group}
    \begin{algorithmic}[1]
	\State \begin{varwidth}[t]{\linewidth}
	    \textbf{Input:}
		$p$ is the current process id
		$P$ is the total number of processes to communicate \par
		$k$ is the number of colors \par
		$G_p(V,E)$ is partition of the input graph at process $p$ \par
		$T_i$ is the $i$th subtemplate to compute \par
		$S_i$ is subset of $S=\{1,\ldots, k\}$ for $i$th subtemplate \par
		$threadIdx$ is the thread Id in process \par
		\textbf{Output:} $C_p(v,T_i,S_i)$ is the updated counts at process $p$ \par
	\end{varwidth}
	\If{($|T_i|$ is large)} \Comment{Adaptive to large $T$}
	\For{$r=p+1,p+2,\dots, P-1, 0, \dots, p-1$}
	\If{threadIdx = 0} \Comment{Communication Pipeline}
	\State Denote $C_{x,y}(v,T_i, S_i)$ as the count value of vertex in process $x$ requested by process $y$  
	\State Compress and send $C_{p,r}(v,T_i, S_i)$ to process $r$ 
	\State Receive and decompress $C_{2p-r,p}(v,T_i,S_i)$ from process $2p-r$
	\Else \Comment{Computation Pipeline}
	\ForAll{$v \in G_p(V,E)$} \Comment{Thread-level parallelism}
	\State \begin{varwidth}[t]{\linewidth}
		Update $C_p(v,T_i,S_i)$ by computing \par 
		received neighbor vertices of $v$ \par 
		from process $2p-r-1$ 
	\end{varwidth}
\EndFor
\EndIf
\EndFor
\Else \Comment{Adaptive to small $T_i$}
\State \begin{varwidth}[t]{\linewidth}
	Process $p$ All-to-All exchanges local counts\par 
	$C_p(v,T_i,S_i)$ with other processes 
\end{varwidth}
\ForAll{$v \in G_p(V,E)$} \Comment{Thread-level parallelism}
\State 
\begin{varwidth}[t]{\linewidth}
	Update $C_p(v,T_i,S_i)$ by computing received \par
	neighbor vertices of $v$ from other processes
\end{varwidth}
\EndFor
\EndIf
\end{algorithmic}
\end{algorithm}

\subsubsection{Pipeline Design}
\begin{figure}[!ht]
    \centering
    \includegraphics[width=0.8\linewidth]{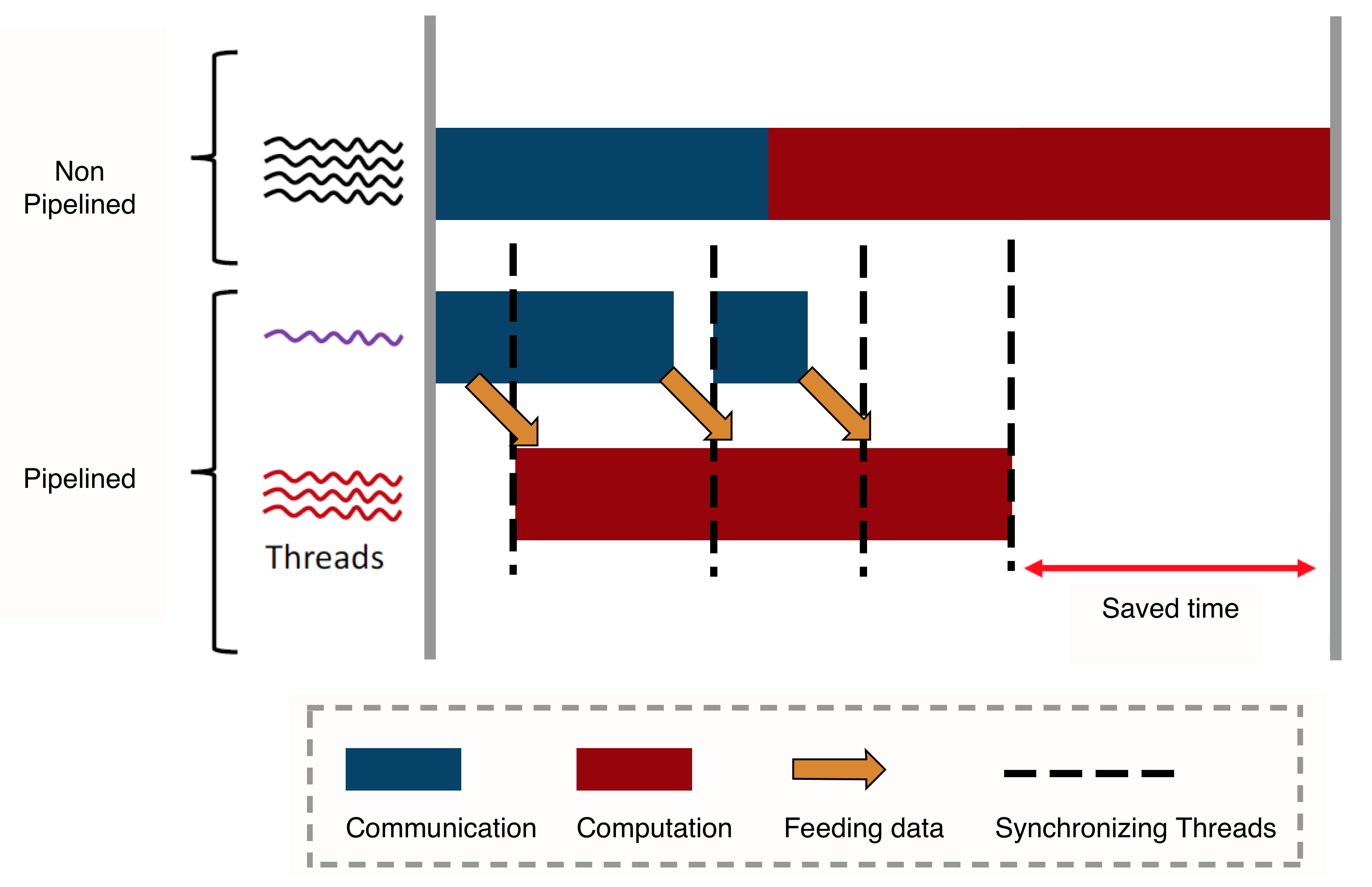}
    \caption{Pipelined Adaptive-Group communication}
    \label{fig:pipeline-design}
\end{figure}
When adding up all $W$ steps in Adaptive-Group, we apply a pipeline design shown in Figure~\ref{fig:pipeline-design}, which includes 
a computation pipeline (red) and a communication pipeline (blue). Given an Adaptive-Group communication in $W$ steps, each pipeline follows 
$W+1$ stages to finish all the work. The first stage is a cold start, where no previous received data exists in the computation pipeline and only the 
communication pipeline is transferring data. For the following $W$ stage, the work in the communication pipeline can be interleaved by the work in the computation pipeline. This interleaving can be achieved by using a multi-threading programming model, where a single thread is in charge of 
the communication pipeline and the other threads are assigned to the computation pipeline 
(see Algorithm~\ref{alg:adaptive-group} line 5 to 13). Since at each stage the computation pipeline 
relies on the data received at the previous stage of the communication pipeline, a synchronization of two pipelines at the end of each stage is required (shown as a dashed line in Figure~\ref{fig:pipeline-design}). The additional performance brought by pipeline 
depends on the ratio of overlapping computation and communication in each stage of two pipelines. We will estimate the bounds of
computation and communication in pipeline design for large templates through an analysis of complexity.
\subsubsection{Complexity Analysis}
When computing subtree $T_i$, we estimate the computation complexity on remote neighbors at step $w$ as:
\begin{equation}
    Comp_{w,p} = O\left({k\choose|T_i|}{|T_i|\choose |T_i'|}\sum_{v\in V_p}N_{r,w}\left(v\right)\right) 
    \label{eq:comp_wp1}
\end{equation}
where $k$ is the number of colors, 
$|T_i|$ is the size of subtree $T_i$ in template $T$, and $T'_i$ is a subtree partitioned from $T_i$
according to Algorithm~\ref{alg:sequential}.\par 

We divide the neighbors of $v$ into local neighbors $N_{l}(v)$ and remote neighbors $N_{r}(v)$.
The $N_{r}(v)$ is made up of neighbors received in each step, $N_{r}(v) = \sum_{w=1}^{W}N_{r,w}(v)$. 
With the assumption of random partitioning $G(V,E)$ by vertices across $P$ processes,
\begin{align}
    E[N_{r,w}(V_p)] &= E[\sum_{v\in V_p} N_{r,w}(v)] \nonumber\\
           &= \sum_{v\in V} E[N_{r,w}(v)]\Pr[v\in V_p] \nonumber \\
           &= \sum_{(u,v)\in E} \Pr[v\in V_p, u\in N_{r,w}(v)] = |E|/P^2
\end{align}
where $|E|$ is the edge number. Further by applying Chernoff bound, we have $N_{r,w}(V_p) = \Theta(|E|/P^2)$ with probability at least $1-1/n^2$.

Therefore, we get the bound of computation as
\begin{align}
    Comp_{w,p} = {k\choose|T_i|}{|T_i|\choose |T'_i|}\Theta\left(N_{r,w}\left(V_p\right)\right)
    = \Theta \left({k\choose |T_i|}{|T_i|\choose |T'_i|}|E|/P^2\right)
    \label{eq:comp_wp}
\end{align}

Similarly, the expectation of peak memory utilization at step $w$ is 
\begin{align}
    PeakMem_{w,p} &= O\left(\sum_{v\in V_p}\Big[C\left(v,T_i\right) + \sum_{u\in N_{r,w}\left(v\right)}C\left(u,T_i\right)\Big]\right) \nonumber\\
    &= O\left({k\choose |T_i|} \left(|V|/P + |E|/P^2\right)\right)
\end{align}
where $C(u,T_i)$ is the length of array (memory space) that holds the combination of color counts for each $u$, and its complexity is bounded by $O({k\choose |T_i|})$ (refer to line 8 of Algorithm~\ref{alg:sequential}).

The communication complexity at step $w$ by Hockney model~\cite{hockney_communication_1994} is 
\begin{align}
    Com_{w,p} &= O\left( \alpha + \delta_{w,p} + \beta\sum_{v\in V_p}\sum_{u\in N_{r,w}\left(v\right)}C\left(u,T_i\right)\right) \nonumber\\
    &= O\left( \alpha + \delta_{w,p} + \beta{k\choose |T_i|}|E|/P^2\right)
\end{align}
where $\alpha$ is the latency associated to the operations in step $w$, $\beta$ is the data transfer time per byte, and 
$\delta_{w,p}$ is the time spent by process $p$ in waiting for other processes because 
of the load imbalance among $P$ processes at step $w$, which is bounded by
\begin{align}
    \delta_{w,p} &= O\left(Max_{q\neq p}\left(Time_{w-1,q}-Time_{w-1,p}\right)\right) \nonumber \\
    &= O\left(Max_{q\neq p}\left(Time_{w-1,q}\right)\right)
    \label{eq:delta}
\end{align}
where $Time_{w-1,q}$ is the execution time of process $q$ at step $w-1$ which is expressed as \par
\begin{equation}
Time_{w-1,q} = Max\left(Comp_{w-1,q}, Com_{w-1,q}\right)
\label{eq:timewq}
\end{equation}
When it comes to the total complexity of all $W$ steps, we assume a routing algorithm described 
in Figure~\ref{fig:ring-order} is used, where $W = P -1$. We obtain the bound for computation as
\begin{align}
    Comp^{pip}_{total,p} &= \sum_{w=1}^{W}Comp_{w,p} \nonumber \\
    &= \Theta \left({k\choose |T_i|}{|T_i|\choose |T'_i|}|E|\left(P-1\right)/P^2\right)
\end{align}
While the peak memory utilization is 
\begin{align}
    PeakMem^{pip}_{total,p} &= O\left(Max_{w}\left(PeakMem_{w,p}\right)\right) \nonumber \\
    &= O\left({k\choose |T_i|} \left(|V|/P + |E|/P^2\right)\right)
    \label{eq:peakm}
\end{align}
The total communication overhead in the pipeline design of all steps $W$ is calculated by
\begin{align}
     Com^{pip}_{total,p} = Com_{w=1,p} + \sum_{w=2}^{W}\left(1-\rho_{w}\right)Com_{w,p}
    \label{eq:totalAGCom}
\end{align}
where $\rho_{w}$ is defined as the ratio of effectively overlapped communication time 
by computation in a pipeline step $w$
\begin{equation}
    \rho_{w} = {Min\left(Comp_{w-1, p}, Com_{w,p}\right)\over Com_{w,p} }, (w>1)
\end{equation}
As the computation per neighbor $u\in N_{r,w}(v)$ for $T_i$ is bounded by ${k\choose |T_i|}{|T_i|\choose |T_i'|}$ and 
communication data volume per $u$ bounded by its memory space complexity ${k\choose |T_i|}$, $Comp_{w,p}$ increases faster than 
$Com_{w,p}$ with respect to the template size $|T_i|$. Therefore, for large templates,  
the computation term $Comp_{w,p}$ is generally larger than the 
communication overhead $Com_{w,p}$ at each step, and we have $\rho_{w} \approx 1$.
Equation~\ref{eq:totalAGCom} is bounded by 
\begin{align}
    Com^{large T, pip}_{total, p} &= O\left(Com_{w=1,p}\right) \nonumber \\ 
    &= O\left(\alpha + \delta_{w=1,p} + {\beta \over P}{k\choose |T_i|}|E|/P^2\right)
    \label{eq:largeTpip}
\end{align}
With large $|T_i|$, we have 
\begin{align}
  \delta_{w=1,p} &= O\left(Max_{q\neq p}\left(Comp_{w=1,q}\right)\right) \nonumber \\
  &= O\left({1\over P^2}{k\choose|T_i|}{|T_i|\choose |T'_i|}|E|\right) 
\end{align}
which is inversely proportional to $P^2$.
The third term in Equation~\ref{eq:largeTpip} is also inversely proportional to $P$. Therefore 
$Com^{largeT,pip}_{total,p}$ shall decrease with an increasing $P$, which implies that the algorithm is scalable with large templates by bounding the communication overhead. \par 
For small templates, there is usually no sufficient workload to interleave communication overhead, which gives a relatively small 
$\rho_{w}$ value in Equation~\ref{eq:totalAGCom} and compromises the effectiveness of pipeline interleaving.  
Even worse, as the transferred data at each step $w$ is small, it cannot effectively leverage the bandwidth of interconnect 
when compared to the all-to-all operation. In such cases, the Adaptive-Group is able to switch back to all-to-all mode and 
ensure a good performance.

\subsubsection{Implementation}
We implement the pipelined Adaptive-Group communication with Harp, where a mapper plays the role of a parallel process, and
mappers can complete various collective communications that are optimized for big data problems. 
In implementation like MPI\_Alltoall, each process $p$ out of $P$ prepares a slot Slot(q) for any other process $q$ that 
it communicates with, and pushes data required by $q$ to Slot(q) prior to the collective communication. The ID label of 
sender and receiver are attached to the slots in a static way, and the program must choose a type of collective operation 
(e.g., all-to-all, allgather) in the compilation stage. \par
\begin{figure}[!ht]
    \centering
    \includegraphics[width=0.8\linewidth]{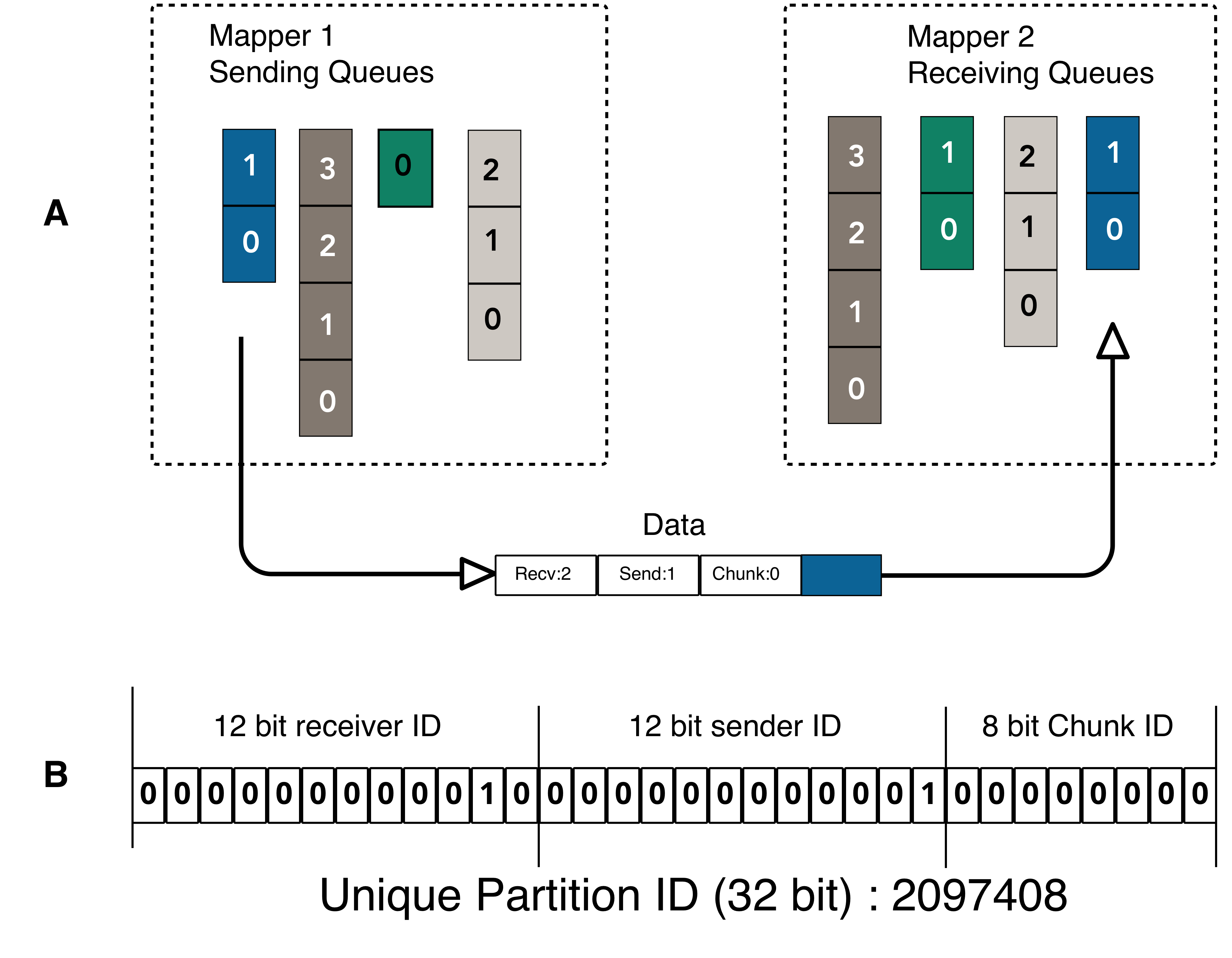}
    \caption{Adaptive-Group tags each data packet with a meta ID, which is used by a routing algorithm for data transfer. 
    Both the meta ID and the routing algorithm are re-configurable on-the-fly}
    \label{fig:tag-system}
\end{figure}
In contrast, each Harp mapper keeps a sender queue and a receiving queue, and each data packet is 
labeled by a meta ID as shown in Figure~\ref{fig:tag-system}. For Adaptive-Group, the meta ID for each packet consists of
three parts (bit-wise packed to a 32-bit integer): the sender mapper ID, the receiver mapper ID, and the offset position 
in the sending queue. A user-defined routing algorithm then decodes the meta ID and delivers the packet in a dynamically-configurable way. The routing algorithm is able to detect template and workload sizes, and switch on-the-fly between pipeline and all-to-all modes.   

\subsection{Fine-grained Load Balance}
\label{sec:neighbor_list_partition}

For an input graph with a high skewness in out-degree of vertex, color-coding imposes a load imbalance issue
at the thread-level. In Algorithm~\ref{alg:sequential} and ~\ref{alg:distriColorCoding}, the task of computing the counts of a certain vertex by looping all entries of its neighbor list is assigned to a single thread. If the max degree of an input graph is several orders of magnitude larger than the average degree, one thread may take orders of magnitude more workload than average. For large templates, this imbalance is amplified by the exponential increase of computing counts for a single vertex in line 9 of Algorithm~\ref{alg:sequential}. \par

To address the issue of workload skewness, we propose a neighbor list partitioning technique, which is implemented by the multi-thread programming library OpenMP. Algorithm~\ref{alg:nbrsplit} illustrates the process of creating the fine-grained tasks assigned to threads.
Given maximal task size $s$, the process detects the neighbor list length $n$ of a vertex $v$. 
If $n$ is beyond $s$, it extracts a sub-list sized $s$ out of the $n$ neighbors and creates a task including neighbors in the sub-list associated to vertex $v$. The same process applies to the remaining part of the truncated list until all neighbors are partitioned. 
If $n$ is already smaller than $s$, it creates a task with all the $n$ neighbors associated to vertex $v$.

\begin{algorithm}[!ht]
	\caption{Create parallel tasks via neighbor list partitioning}
	\label{alg:nbrsplit}
	\begin{algorithmic}[1]
		\State \begin{varwidth}[t]{\linewidth}
		\textbf{Input:} $s$ is user-defined maximal task size \par
		$V$ is local vertices \par
		$N_v$ is neighbor list of $v\in V$ \par
		$n$ is the number of neighbors \par 
		$l$ is the length of new task \par
		$pos$ is the offset of sub-list \par
		\textbf{Output:} $Q$ stores created tasks \par
		\end{varwidth}
		\ForAll{$v \in V$}
		\If{$|N_v| < s$}
		\State $Q$ add task(v, $N_v$)
		\Else
		\State $n \gets |N_v|$
		\State $pos \gets 0$
		\While{$n > 0$}
		\State $l \gets Min(n, s)$
		\State $Q$ add Task(v, $N_v(pos:pos+l)$)
		\State $pos += l$
		\State $n -= l$ 
	\EndWhile
	\EndIf
	\EndFor
	\State shuffle tasks in $Q$
\end{algorithmic}
\end{algorithm}
The neighbor list partitioning ensures that no extremely large task is assigned to a thread by bounding the task size to $s$, which
improves the workload balance at thread-level. However, it comes with a race condition if two threads are updating tasks associated to the 
same vertex $v$. We use atomic operations of OpenMP to resolve the race condition and shuffle the created task queue at 
line 17 of Algorithm~\ref{alg:nbrsplit} to mitigate the chance of conflict.  

\section{Evaluation of Performance and Analysis of Results}
\label{sec:eval}

\subsection{Experimentation Setup}
\label{sec:exp}
We conduct a set of experiments by implementing 4 code versions of distributed color-coding algorithm with Harp-DAAL: Naive, Pipeline, Adaptive and AdaptiveLB (Load Balance). Table~\ref{tab:hd-version} lists individual optimization technique for experiments. They aim to systematically investigate the impact of each optimization, which addresses the sparse irregularity, the low computation to communication ratio or the high memory footprint issues of subgraph counting. \\

\begin{table}[!ht]
	\tiny
  \caption{Harp-DAAL implementations in experiments}
  \label{tab:hd-version}
  \begin{tabular}{>{\raggedright}p{2.0cm}>{\raggedright}p{2.0cm}>{\raggedright}p{2.0cm}>{\raggedright}p{1.5cm}}
    \toprule
    Implementation & Communication Mode & Adaptive Switch & Neighbor list partitioning \\
    \midrule
    Naive & all-to-all & Off & Off \\
    Pipeline & Pipeline & Off & Off \\
    Adaptive & all-to-all/pipeline & On & Off \\
    AdaptiveLB & all-to-all/pipeline & On & On \\
  \bottomrule
\end{tabular}
\end{table}

\begin{table}[!h]
	\caption{Datasets in experiments (K=$10^3$, M=$10^6$,B=$10^9$)}
	\label{tab:datasets}
\resizebox{\linewidth}{!}{%
	\begin{tabular}{lllllll}
		\toprule
		Data  &  Vertices &  Edges &  Avg Deg &  Max Deg & Source & Abbreviation\\ 
		\midrule
		Miami &  2.1M & 51M &  49 &  9868 & social network & MI\\ 
		Orkut &  3M & 230M  & 76 & 33K & social network & OR\\ 
		NYC   & 18M & 480M  & 54 & 429 & social network & NY\\ 
		Twitter & 44M & 2B & 50 & 3M &  Twitter users & TW\\ 
		Sk-2005 & 50M & 3.8B & 73 & 8M &  UbiCrawler & SK\\ 
		Friendster & 66M & 5B & 57 & 5214 & social network & FR\\ 
		RMAT-250M(k=1,3,8) & 5M & 250M & 100,102,217 & 170,40K,433K & PaRMAT & R250K1,3,8\\
		RMAT-500M(k=3) & 5M & 500M & 202 & 75K & PaRMAT & R500K3\\
		\bottomrule
	\end{tabular}
}
\end{table}

\begin{figure}[!ht]
  \centering
  \includegraphics[width=0.6\linewidth]{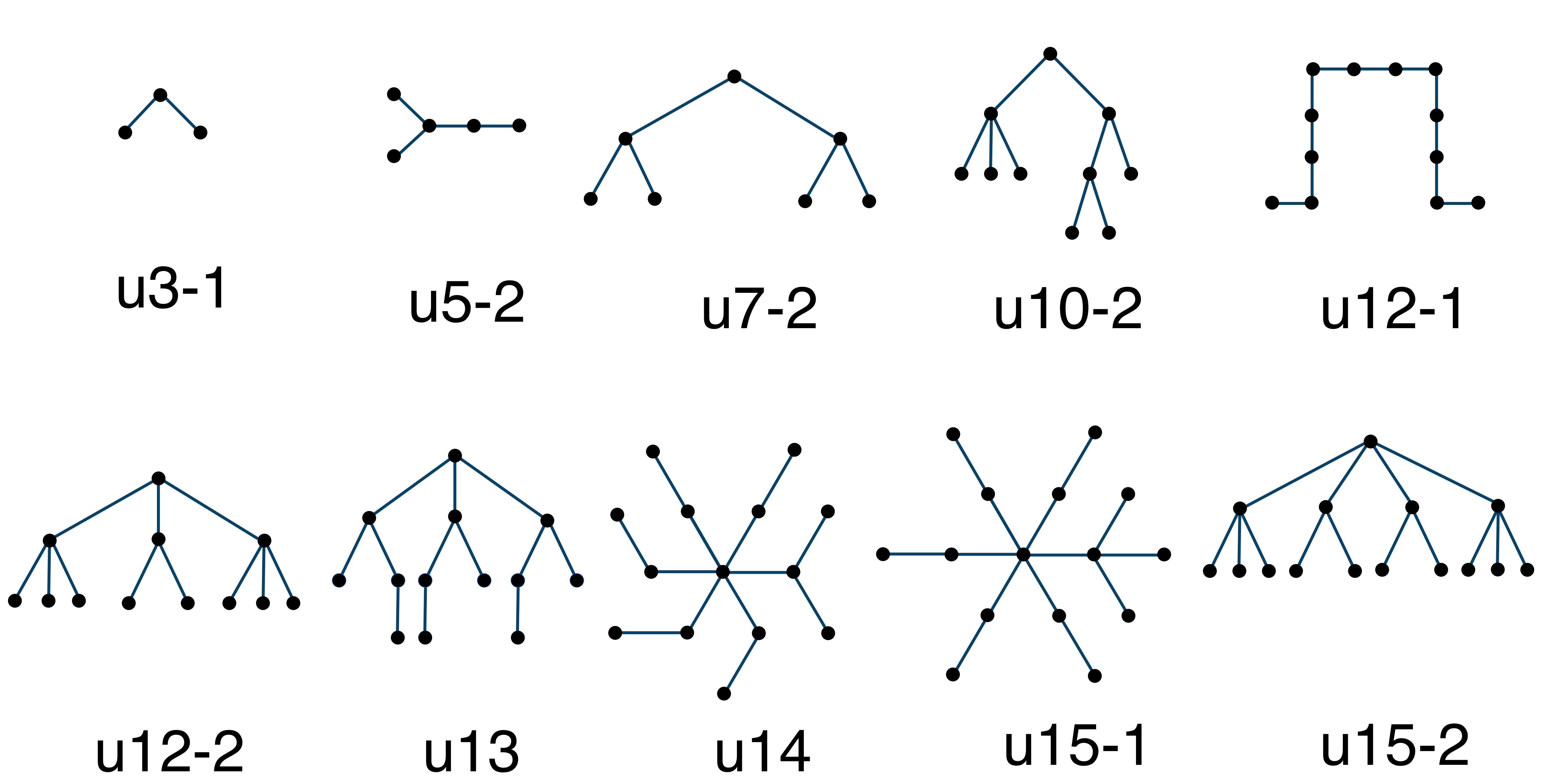}
  \caption{Tree templates used in experimentation with growing sizes and different shapes} 
  \label{fig:bakTemplates}
\end{figure}
\begin{table}[!ht]
    \caption{Computation intensity of templates}
    \label{tab:tpcomplexity}
\tiny
    \begin{tabular}{>{\raggedright}p{1.5cm}>{\raggedright}p{1.5cm}>{\raggedright}p{1.5cm}>{\raggedright}p{1.5cm}}
    \toprule
    Template & Memory Complexity & Computation Complexity & Computation Intensity  \\
    \midrule
         u3-1  & 3 & 6 & 2 \\
         u5-2  & 25 & 70 & 2.8 \\
         u7-2  & 147 & 434 & 2.9 \\
         u10-2 & 1047 & 5610 & 5.3 \\
         u12-1 & 4082 & 24552 & 6.0 \\
         u12-2 & 3135 & 38016 & 12 \\
         u13 & 4823 & 109603 & 22 \\
         u14 & 7371 & 242515 & 32 \\
         u15-1 & 12383 & 753375 & 60 \\
         u15-2 & 15773 & 617820 & 39 \\
    \bottomrule
    \end{tabular}
\end{table}

We use synthetic and real datasets in our experiments which are summarized in Table~\ref{tab:datasets}. Miami, Orkut~\cite{barrett_generation_2009}\cite{snapnets}\cite{yang_defining_2012}, Twitter~\cite{cha_measuring_2010}, SK-2005~\cite{davis_university_2011}, and Friendster~\cite{snapnets} are datasets generated by real applications. RMAT synthetic datasets are generated by the RMAT model~\cite{chakrabarti_r-mat:_2004} by specifying the size and skewness. 
Specifying a higher skewness generates a highly imbalanced distribution of out-degree for input graph datasets. Therefore, we can use different skewness of RMAT datasets to study the impact of unbalanced workload on the performance. 
The different sizes and structures of the tree templates used in the experiments are shown in Figure~\ref{fig:bakTemplates}, where 
templates from u3-1 to u12-2 are collected from ~\cite{slota_parallel_2015}, while u13 to u15 are the largest tree subgraphs being tested to date. \par
We observe that the size and shape of sub-templates affect the ratio of computation and communication in our experiments. This corresponds to code line 8 
of Algorithm~\ref{alg:sequential}, where each sub-template $T_i$ is partitioned into trees $T'_i$ and $T''_i$. 
The space complexity for each neighbor $u\in N(v)$ is bounded by ${k\choose |T_i|}$ when computing sub-template $T_i$, and is proportional to the communication data volume. The computation, which depends on the shape of the template, is bounded by ${k\choose |T_i|}{|T_i|\choose |T'_i|}$. In Table~\ref{tab:tpcomplexity}, the memory space complexity is denoted as $\sum_{i}{k\choose |T_i|}$, and the computation complexity is $\sum_{i}{k\choose |T_i|}{|T_i|\choose |T'_i|}$. In this chapter, we define the \textit{computation intensity} as the ratio of computation versus communication (or space) for a template in Figure~\ref{fig:bakTemplates}. 
For example, the computation intensity generally increases along with the template size from u3-1 to u15-2. However, for the same template size, template u12-2 has a computation intensity of 12 while u12-1 only has 6. We will use these definitions and refer to their values when analyzing the experiment results in the rest of sections.\par
All experiments run on an Intel Xeon E5 cluster with 25 nodes. Each node is equipped with two sockets of Xeon 
E5 2670v3 (2$\times$12 cores), and 120 GB of DDR4 memory. We use all 48 threads by default in our tests, and InfiniBand is enabled in either Harp or the MPI communication library. 
Our Harp-DAAL codes are compiled by JDK 8.0 and Intel ICC Compiler 2016 as recommended by Intel. The MPI-Fascia~\cite{slota_parallel_2015} codes are compiled by OpenMPI 1.8.1 as recommended by its developers.

\subsection{Scaling by Adaptive-Group Communication}
\label{sec:scaling_adaptive}
\begin{figure}[!ht]
    \centering
    \includegraphics[width=0.6\linewidth]{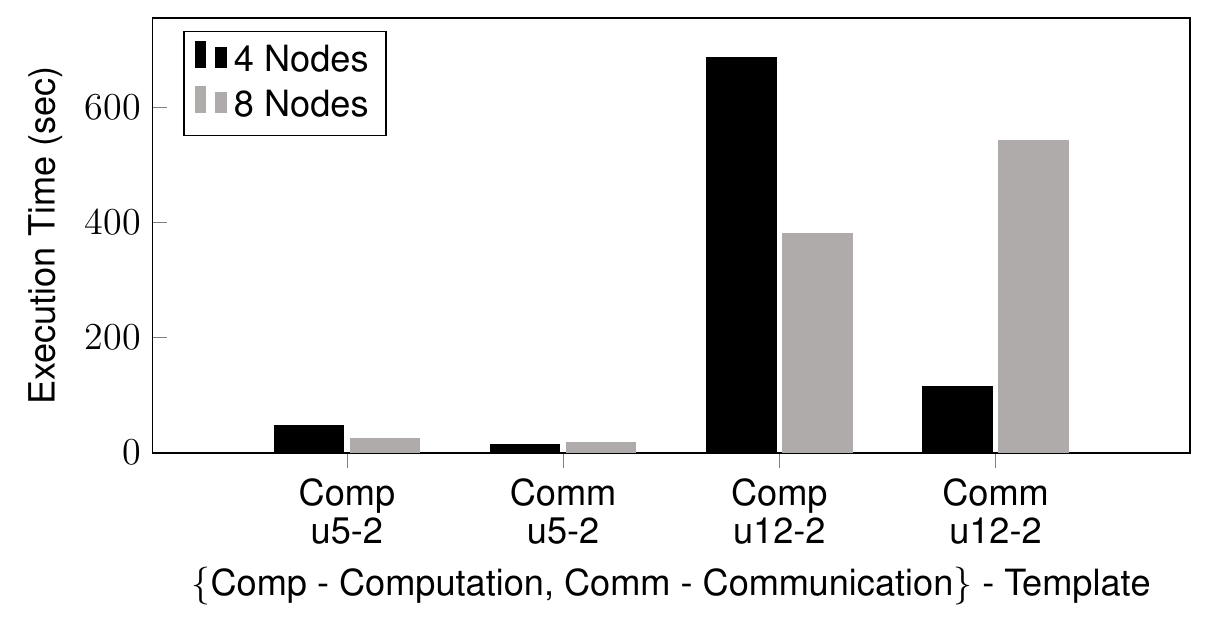}
    \caption{Scaling up template sizes on dataset R500K3 for Harp-DAAL Naive implementation from 4 cluster nodes to 8 cluster nodes}
    \label{fig:naive-challenge}
\end{figure}
\begin{figure}[!t]
    \centering
    \includegraphics[width=1.0\linewidth]{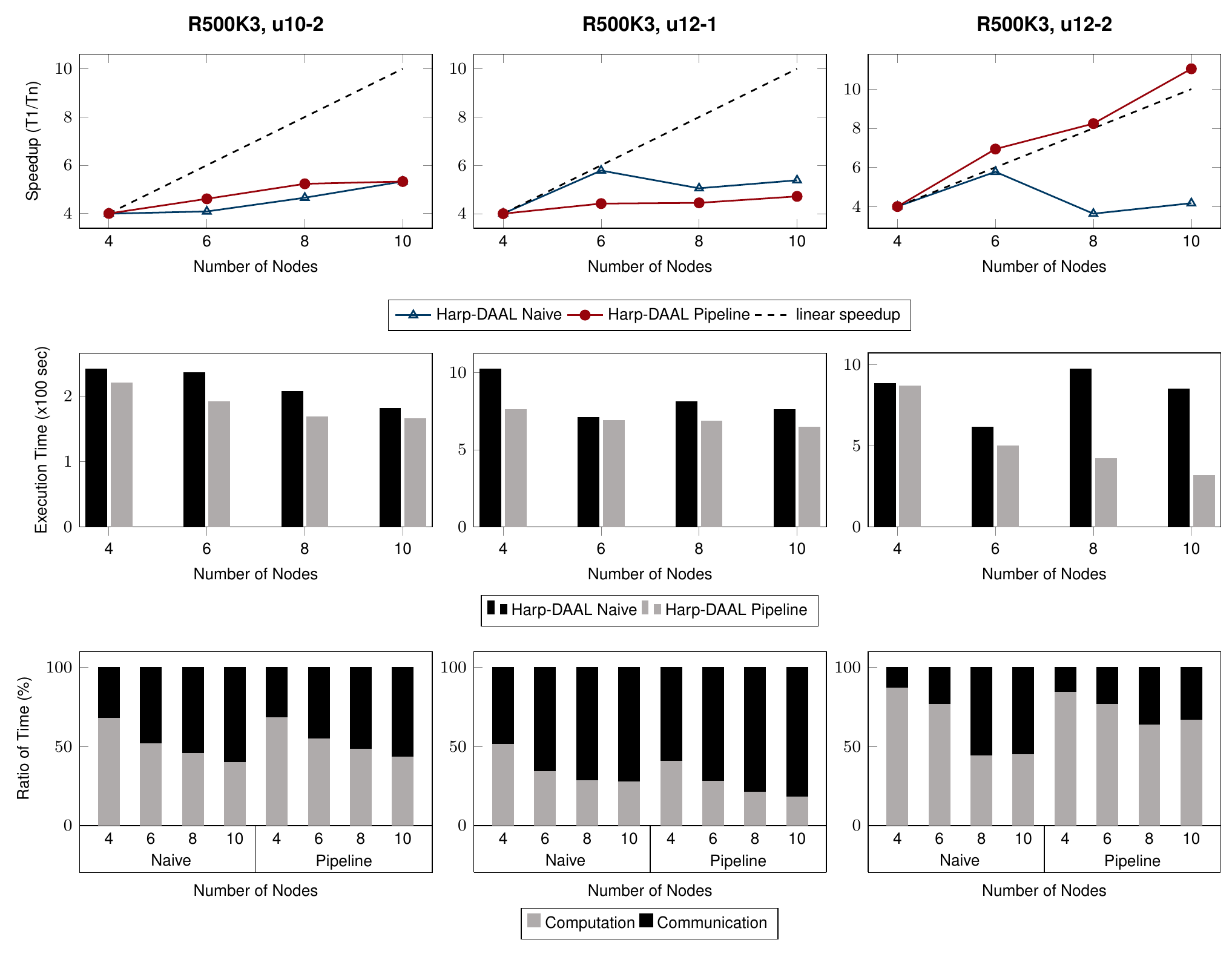}
    \caption{Strong scaling tests on dataset R500K3 from 4 to 10 cluster nodes with large templates (u10-2, u12-1, u12-2). 
    First row gives the speedup starting from 4 cluster nodes since a single node cannot hold the dataset; the second row compares the total execution time from two implementations; the third row is the ratio of compute/communicate time in the total execution time.}
    \label{fig:naive-pipeline-sscaling}
\end{figure}
\begin{figure}[!t]
    \centering
    \hspace{-0.5cm}
    \includegraphics[width=0.8\linewidth]{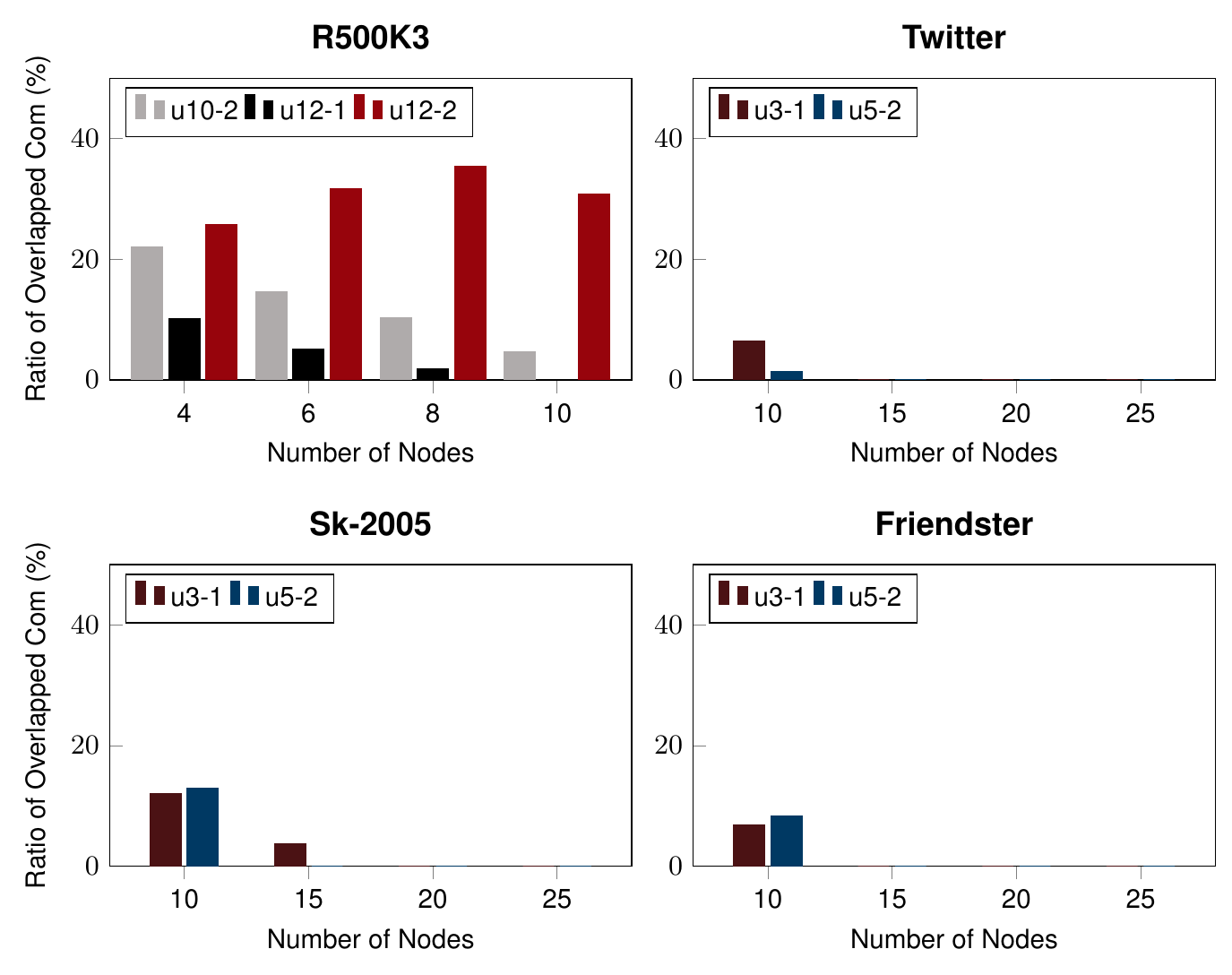}
    \caption{The ratio $\rho$ of overlapped communication/total communication by Harp-DAAL Pipeline, tests on R500K3 for large templates (u10-2, u12-1, u12-2), 
    and Twitter, Sk-2005, Friendster for small templates u3-1, u5-2}
    \label{fig:pipeline-overlap-ratio}
\end{figure}
\begin{figure}[!t]
    \centering
    \includegraphics[width=\linewidth]{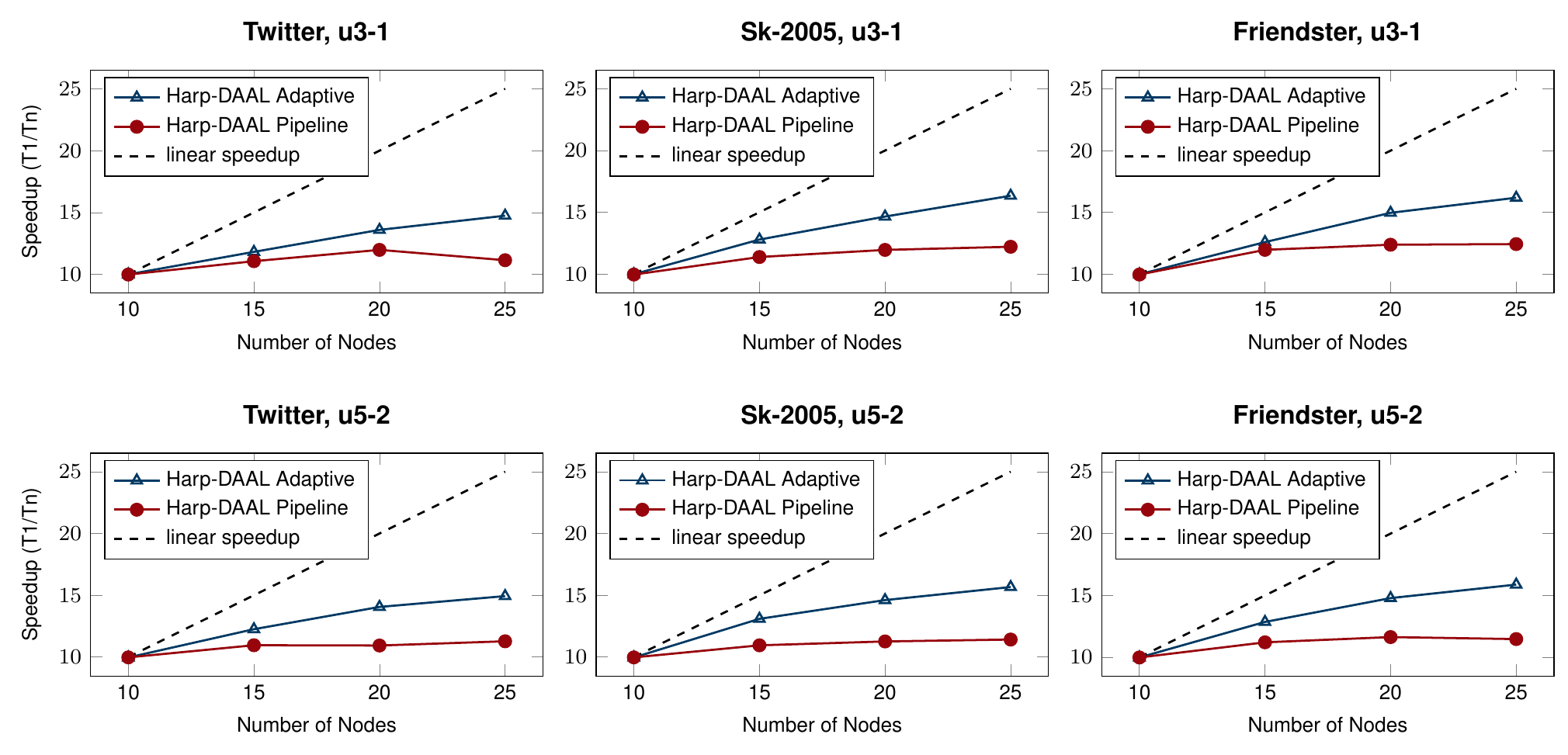}
    \caption{Strong scaling tests on large dataset Twitter, SK-2005, Friendster from 10 
    cluster nodes to 25 cluster nodes with small templates (u3-1, u5-2). Harp-DAAL Adaptive switches to all-to-all mode and outperforms pipeline.}
    \label{fig:naive-pipeline-sscaling-smallT}
\end{figure}
\begin{figure}[htbp]
    \centering
    \includegraphics[width=0.6\linewidth]{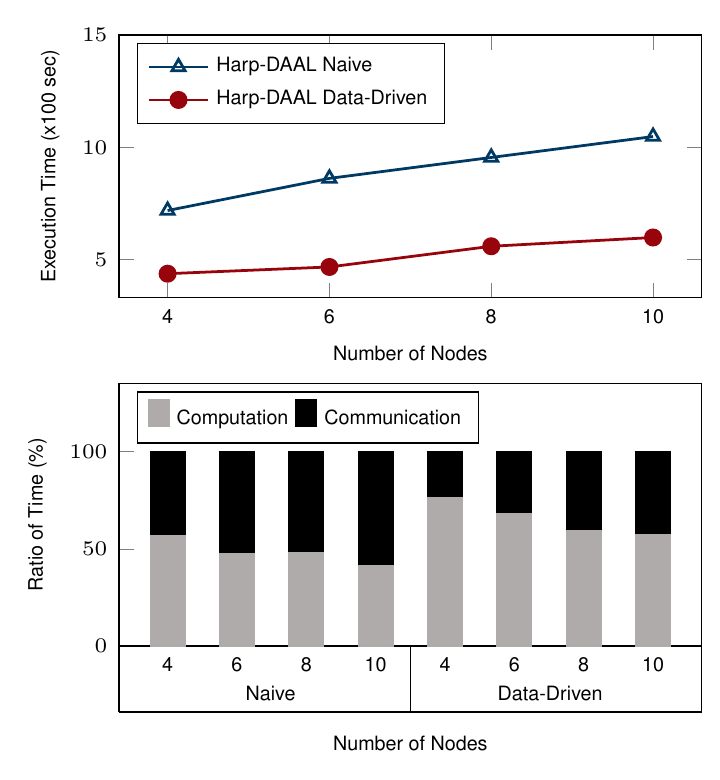}
    \caption{Weak scaling of RMAT with skeweness 3; the workload is proportional to the cluster nodes: e.g., 5 million vertices with 250 million edges on 4 
    cluster nodes, and 7.5 million vertices with 375 million edges on 6 cluster nodes. }
    \label{fig:naive-pipeline-weakscale}
\end{figure}
We first conduct a baseline test with the naive implementation of distributed color-coding. When the subgraph template size is scaled up as shown in Figure~\ref{fig:naive-challenge}, 
we have the following observations: 1) For small template u5-2, computation decreases by 2x when scaling from 4 to 8 nodes while communication only increases by 13\%. 
2) For large template u12-2, doubling cluster nodes only reduces computation time by 1.5x but communication grows by 5x. 
It implies that the all-to-all communication within the Naive implementation does not scale well on large templates.\par 

To clarify the effectiveness of the Harp-DAAL Pipeline on large templates, Figure~\ref{fig:naive-pipeline-sscaling} compares strong scaling speedup, total execution time, and ratio of communication/computation time between the
Naive and Pipeline implementation versions on Dataset R500K3, which has skewness similar to real application datasets such as Orkut. For template u10-2, Harp-DAAL Pipeline only slightly outperforms Harp-DAAL Naive in terms of speedup and total execution time. However, for u12-2, 
this performance gap increases to 2.3x (8 nodes) and 2.7x (10 nodes) in 
execution time, and the speedup is significantly improved starting from 8 nodes. The result is consistent with 
Table~\ref{tab:tpcomplexity}, where u12-2 has 2 times higher computation intensity than u10-2, which provides the pipeline design of sufficient 
workload to interleave the communication overhead.
The ratio charts of Figure~\ref{fig:naive-pipeline-sscaling} also confirm this result that Harp-DAAL Pipeline has more than 65\% of computation on 8 and 10 nodes, while the computation ratio for Harp-DAAL Naive is below 50\% when scaling on 8 and 10 nodes.
Although template u12-1 has the same size as template u12-2, it only has half of the computation intensity as shown in Table~\ref{tab:tpcomplexity}. 
According to Equation~\ref{eq:totalAGCom}, the low computation intensity on u12-1 reduces the overlapping ratio $\rho$, and we find in 
Figure~\ref{fig:pipeline-overlap-ratio} that Harp-DAAL Pipeline has less than 10\% of overlapping ratio for u12-1, while   
u12-2 keeps around 30\% when scaling up to 10 cluster nodes. \par

For small templates similar to u3-1 and u5-2 which have low computation intensities, we shall examine the effectiveness 
of adaptability in Harp-DAAL Adaptive, where the code switches to all-to-all mode. 
In Figure~\ref{fig:naive-pipeline-sscaling-smallT}, we did the strong scaling tests with small templates u3-1 and u5-2. Results show that when compared to Harp-DAAL Pipeline, Harp-DAAL Adaptive has a better speedup for tests of both u3-1 and u5-2 on 
three large datasets: Twitter, Sk-2005, and Friendster. Also, the poor performance of Harp-DAAL Pipeline is due to the
low overlapping ratio in Figure~\ref{fig:pipeline-overlap-ratio} for Twitter, Sk-2005, and Friendster, where $\rho$ drops to near zero quickly after scaling to more than 15 nodes. \par 

In addition to strong scaling, we present weak scaling tests in Figure~\ref{fig:naive-pipeline-weakscale} for template u12-2. We generate a group of RMAT datasets with skewness 3 and an increasing number of vertices and edges proportional to the running cluster nodes. By fixing the workload on each cluster node, the weak scaling on the Harp-DAAL Pipeline reflects the additional communication overhead when more cluster nodes are used. For the Harp-DAAL Pipeline, execution time grows only by 20\% with cluster nodes growing by 2 (from 4 nodes to 8 nodes). From the ratio chart in Figure~\ref{fig:naive-pipeline-weakscale}, it is also clear that the Naive implementation has its communication ratio increased to more than 50\% by using 8 cluster nodes while the communication ratio of Pipeline implementation stays under 40\%. 

\subsection{Fine-grained Load Balance}
\label{sec:imbalance_workload}
\begin{figure}[!ht]
    \centering
    \includegraphics[width=\linewidth]{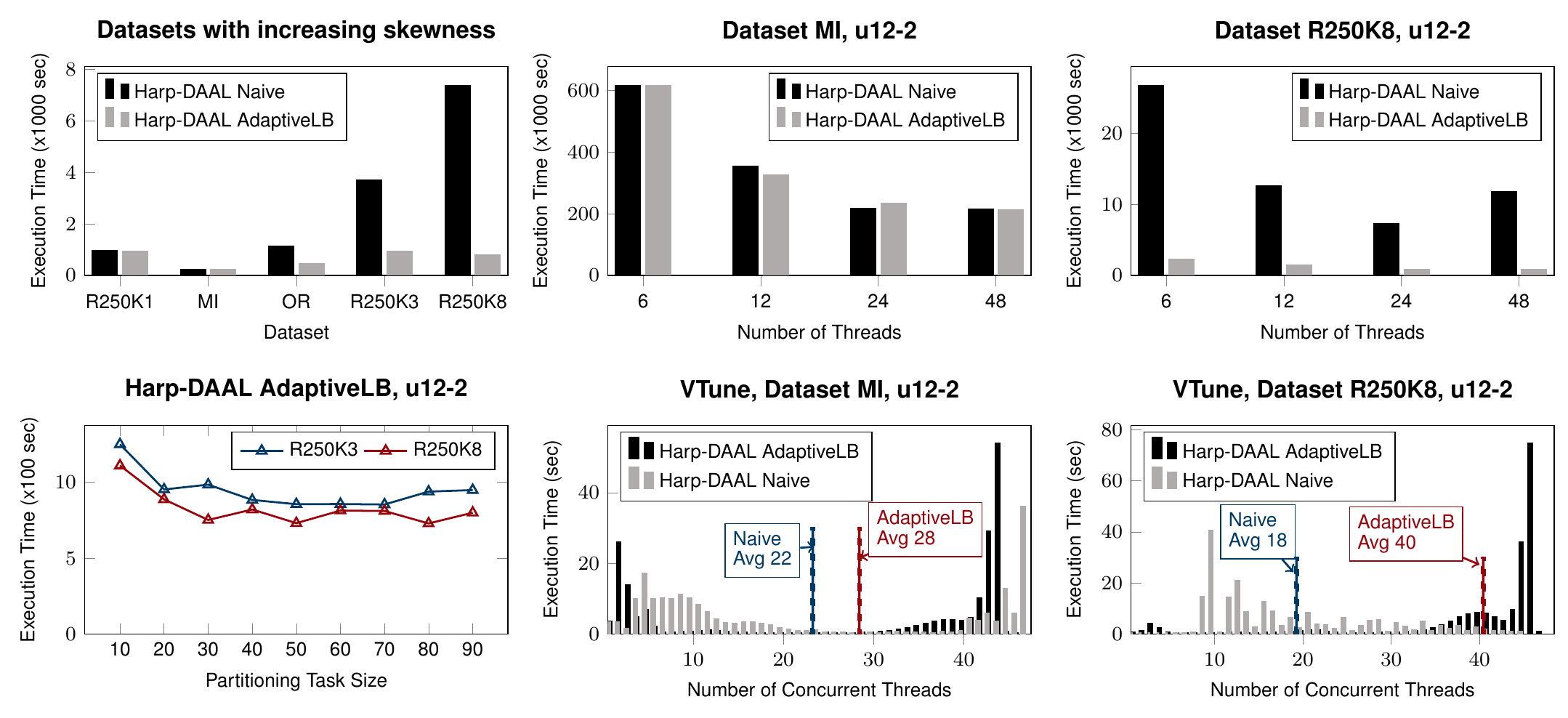}
    \caption{Execution details on a single Xeon E5 node (x2 sockets, and a total of 24 physical cores). The default thread number in test is 48 and partitioned neighbor list is 50.}
    \label{fig:intra-node}
\end{figure}

Although Adaptive-Group communication and pipeline design mitigate the node-level load imbalance caused by skewness of neighbor list length for each vertex in input graph, it can not resolve fine-grained workload imbalance at thread-level 
inside a node. By applying our neighbor list partitioning technique, we compare the performance of Harp-DAAL AdaptiveLB with Harp-DAAL Adaptive on datasets with different skewness. In Figure~\ref{fig:intra-node}, we first compare the datasets with increasing skewness shown in Table~\ref{tab:datasets}. With R250K1 and MI having 
small skewness, the neighbor list partitioning barely gains any advantage, and its benefit starts to appear from dataset OR by 2x improvement of 
the execution time. For a dataset with high skewness such as R250K8 with u12-2 template, this acceleration achieves up to 9x the execution time as shown in Figure 11. \par 
When scaling threads from 6 to 48, for dataset MI having small skewness, the execution time does not improve much. For R250K8, Harp-DAAL AdaptiveLB maintains a good performance compared to Naive implementation. In particular, the thread-level performance of Harp-DAAL Naive drops down after using more than physical core number (24) of 
threads, which implies a suffering from hyper threading. However, Harp-DAAL AdaptiveLB is able to keep the performance 
unaffected by hyper threading. To further justify the thread efficiency of Harp-DAAL AdaptiveLB, we measure the thread concurrency by VTune. 
The histograms show the distribution of execution time by the different numbers of concurrently running threads. For dataset MI, the number of average concurrent
threads of Harp-DAAL Naive and AdaptiveLB are close (22 versus 28) because the dataset MI does not have severe load imbalance caused by skewness.
For dataset R250K8, the number of average concurrent threads of Harp-DAAL AdaptiveLB outperforms that of Harp-DAAL Naive by around 2x (40 versus 18).\par
Finally, we study the granularity of task size and how it affects partitioning of the neighbor list. In Algorithm~\ref{alg:nbrsplit}, each task of updating
neighbor list is bounded by a selected size $s$. If $s$ is too small, there will be a substantial number of created tasks, which adds additional thread scheduling and synchronization overhead. If $s$ is too large, it can not fully exploit the benefits of partitioning neighbor list. There exists a range
of task granularity which can be observed in the experiments on R250K3 and R250K8. To fully leverage the neighbor list partitioning, a task size between 40 and 60 gives better performance than the other values.

\subsection{Peak Memory Utilization}
\label{sec:reduce_peakmem}
\begin{figure}[!ht]
	\centering
	\includegraphics[width=\linewidth]{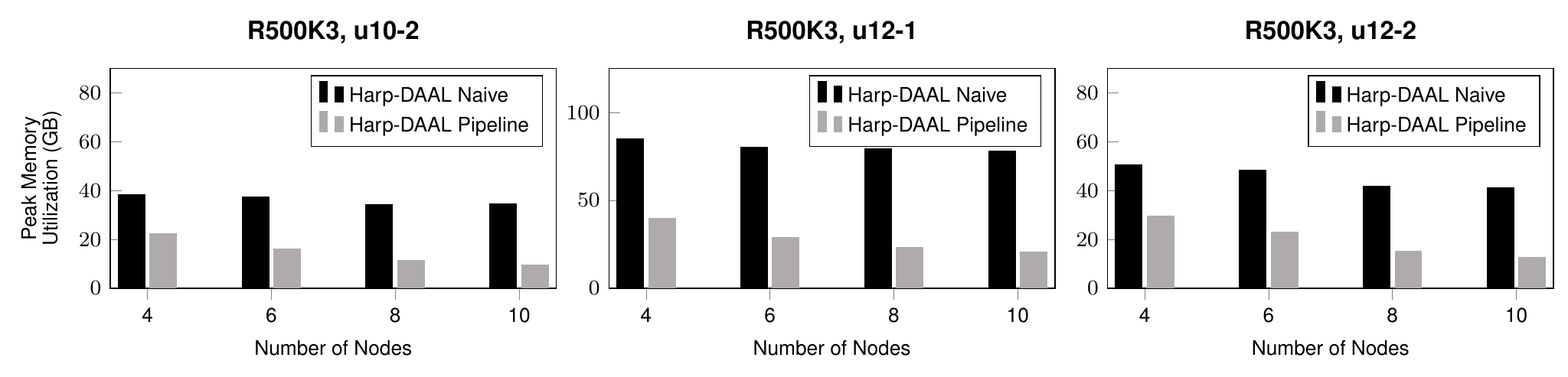}
	\caption{Peak memory utilization for Harp-DAAL Naive and Harp-DAAL Adaptive on dataset R500K3 with templates u10-2,u12-1, u12-2 from 4 to 10 nodes}
	\label{fig:peakmem}
\end{figure}
Adaptive-Group communication and pipeline design also reduce the peak memory utilization at each node. 
According to Equation~\ref{eq:peakm}, peak memory utilization depends on two terms: the $C(v,T)$ from local vertices $V_p$ and 
$C(u,T)$ from remote neighbors $u\in N_{r,w}(v)$. When total $|V|$ of dataset is fixed, $|V_p|$ decreases
with increasing process number $P$ and thus reduces the first peak memory term. 
The second term associated with $u$ at step $w$ is 
also decreasing along with $P$ because more steps ($W=P-1$) leads to small data volume involved in 
each step.  
In Figure~\ref{fig:peakmem}, we observe this reduction of peak memory utilization along with the growing number of cluster nodes from 4 to 10. 
Compared to Harp-DAAL Naive, Harp-DAAL Pipeline reduces the peak memory utilization by 2x on 4 nodes, and this saving grows to around 5x for large templates u10-2, u12-1, and u12-2.

\subsection{Overall Performance}
\label{sub:exp:scaleout}
\begin{figure}[!ht]
	\centering
	\includegraphics[width=\linewidth]{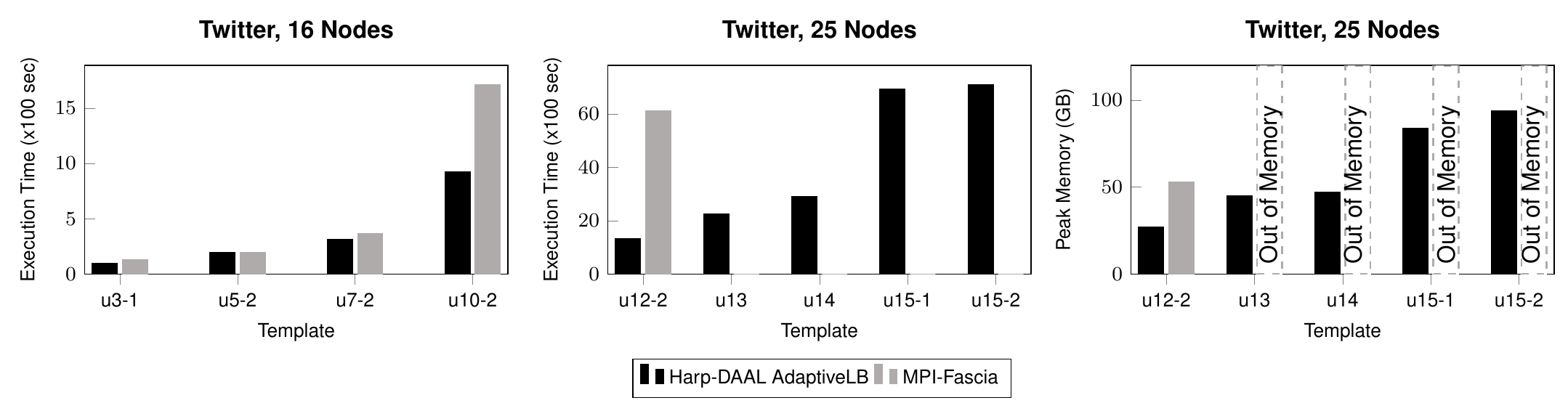}
	\caption{Overall performance of Harp-DAAL AdaptiveLB vs. MPI-Fascia with increasing template sizes from u3-1 to u15-2}
	\label{fig:fullopt-dataset-var}
\end{figure}
\begin{figure}[!t]
    \centering
    \includegraphics[width=0.8\linewidth]{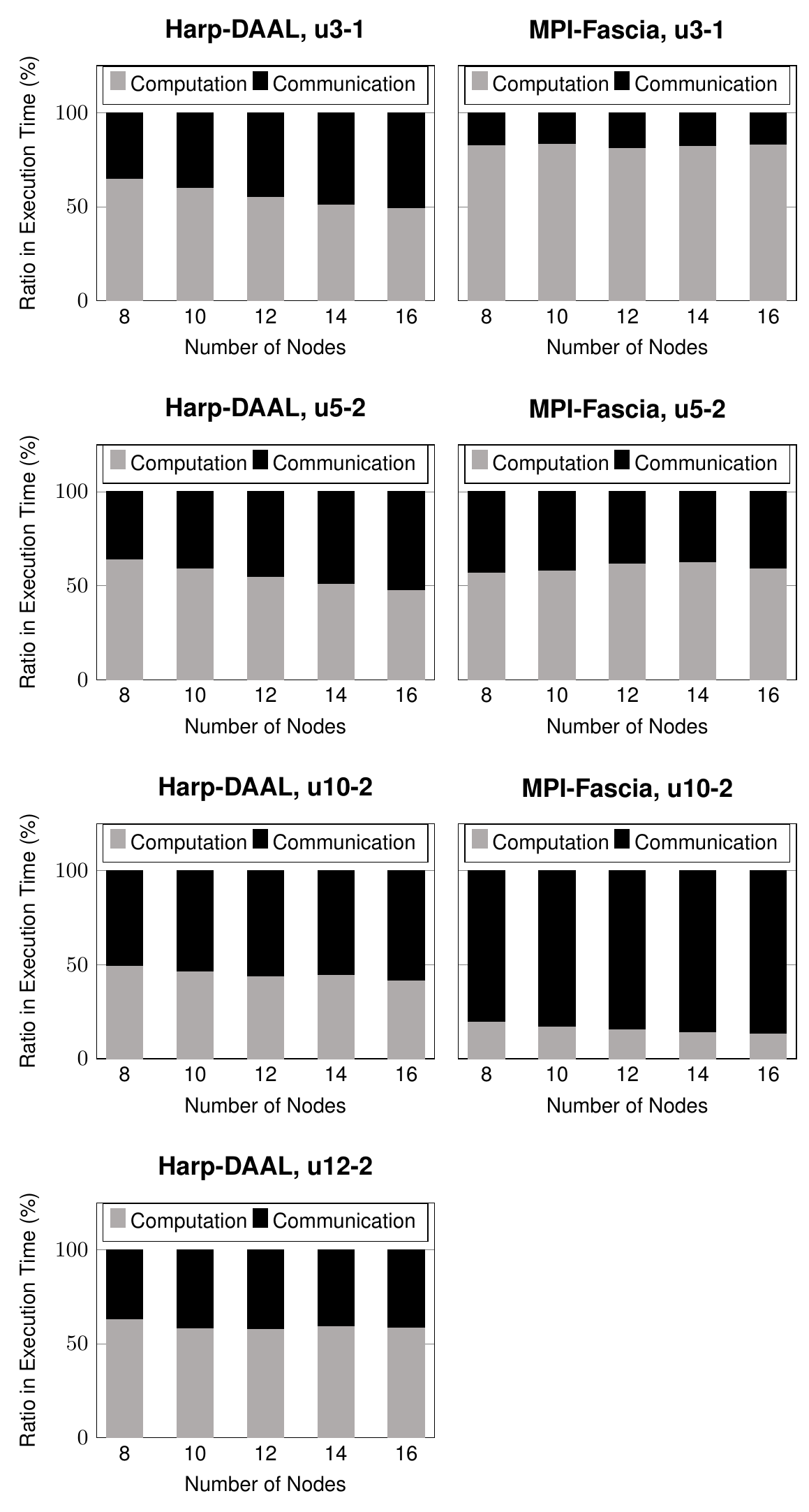}
    \caption{The ratio of computation versus communication in total execution time for Harp-DAAL AdaptiveLB and MPI-Fascia}
    \label{fig:fullopt-sscale-ratio}
\end{figure}
\begin{figure}[!ht]
    \centering
    \includegraphics[width=0.6\linewidth]{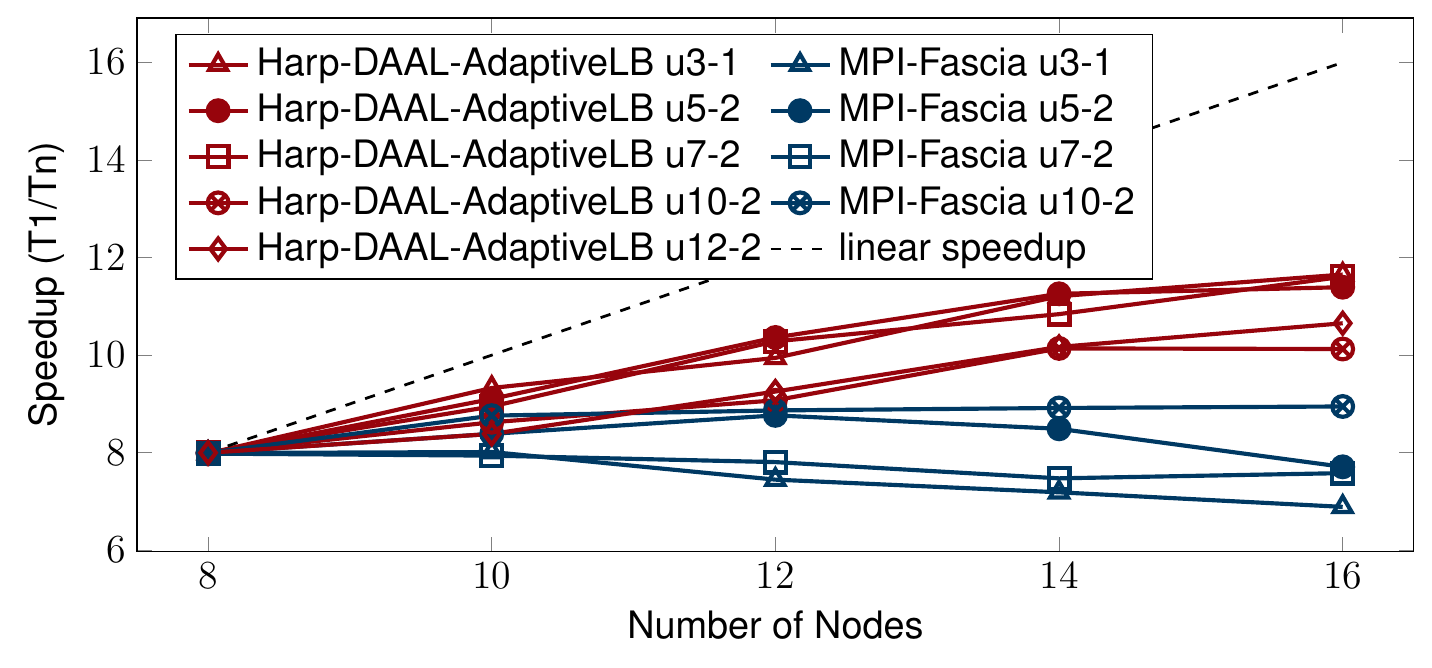}
    \caption{Strong scaling of Harp-DAAL AdaptiveLB vs. MPI-Fascia on Twitter with template sizes from u3-1 to u12-2}
    \label{fig:fullopt-sscale}
\end{figure}
Figure~\ref{fig:fullopt-dataset-var} shows a comparison of Harp-DAAL AdaptiveLB versus MPI-Fasica in total execution time with growing templates on 
Twitter dataset. 
For small templates u3-1, u5-2, and u7-2, Harp-DAAL AdaptiveLB performs comparably or slightly better. Small templates can not fully exploit the efficiency of pipeline due to low computation intensity. 
For large template u10-2, Harp-DAAL AdaptiveLB achieves 2x better performance than MPI-Fascia, and it continues to gain by 5x better performance for u12-2. Beyond u12-2, Harp-DAAL AdaptiveLB can still scale templates up to u15-2. MPI-Fascia can not run templates larger than u12-2 on Twitter because of high peak memory utilization over the 120 GB memory limitation per node. \par 
Figures~\ref{fig:fullopt-sscale-ratio} and \ref{fig:fullopt-sscale} further compare the strong scaling results between Harp-DAAL AdaptiveLB 
and MPI-Fascia. Scaling from 8 nodes to 16 nodes, Harp-DAAL AdaptiveLB achieves better speedup than MPI-Fascia for templates growing from 
u3-1 to u12-2. MPI-Fascia cannot run Twitter on 8 nodes due to its high peak memory utilization. The ratio charts in Figure~\ref{fig:fullopt-sscale-ratio} give more details about the speedup, where MPI-Fascia has a comparable communication overhead ratio in execution 
time for small templates u3-1 and u5-2; however, the communication ratio increases to 80\% at template u10-2 while Harp-DAAL AdaptiveLB 
keeps communication ratio around 50\%. At template u12-2, Harp-DAAL AdaptiveLB further reduces the communication overhead to around 40\% because 
the adaptive-group and pipeline favors large templates with high computation intensity. 

\section{Related Work}
\label{sec:related_work}
Subgraphs of size $k$ with an independent set of size $s$ can be counted in time roughly $O(n^{k-s}\text{poly}(n))$ 
through matrix multiplication based methods \cite{vassilevska2009finding, kowaluk:soda11}. There is substantial work on parallelizing the color-coding technique. ParSE\cite{zhao_subgraph_2010} is the first 
distributed algorithm based on color-coding that scales to graphs with millions of vertices with tree-like template  
size up to 10 and hour-level execution time . SAHAD \cite{zhao_sahad:_2012} expands this algorithm to labeled templates of up to 12 vertices on a graph with 9 million of vertices within less than an hour by using a Hadoop-based implementation. FASCIA \cite{slota_fast_2013, slota_complex_2014, slota_parallel_2015} is the state-of-the-art color-coding treelet counting tool. By highly optimized data structure and MPI+OpenMP implementation, it supports tree template of size up to $10$ vertices in billion-edge networks in a few minutes. Recent work \cite{chakaravarthy_subgraph_2016} also explores the topic of a more complex template with tree width 2, which scales up to $10$ vertices for graphs of up to $2M$ vertices. The original color-coding technique has been extended in various ways, e.g., a derandomized version~\cite{alon:talg10}, and to other kinds of subgraphs. 

\section{Conclusion}
\label{sec:conclusion}

Subgraph counting is a NP-hard problem with many important applications on large networks. We propose a novel pipelined communication scheme for finding and counting large tree templates. The proposed approach simultaneously addresses the sparse irregularity, the low computation to communication ratio and high memory footprint, which are difficult issues for scaling of complex graph algorithms. The methods are aimed at large subgraph cases and use approaches that make the method effective as graph size, subgraph size, and parallelism increase.
Our implementation leverages the Harp-DAAL framework adaptively and improves the scalability by switching the communication modes based on the size of subgraph templates. Fine-grained load balancing is achieved at runtime with thread level parallelism. We demonstrate that our proposed approach is particularly effective on irregular subgraph counting problems and problems with large subgraph templates. For example, it can scale up to the template size of 15 vertices on Twitter datasets (half a billion vertices and 2 billion edges) while achieving 5x speedup over the state-of-art MPI solution. For datasets with high skewness, the performance improves up to 9x in execution time. The peak memory utilization is reduced by a factor of 2 on large templates (12 to 15 vertices) compared to existing work. Another successful application has templates of 12 vertices and a massive input Friendster graph with 0.66 billion vertices and 5 billion edges. All experiments ran on a 25 node cluster of Intel Xeon (Haswell 24 core) processors. Our source code of subgraph counting is available in the public github domain of Harp project\cite{harpdaalweb}.\par

In future work, we can apply this Harp-DAAL subgraph counting approach to other data-intensive irregular graph applications such as random subgraphs and obtain scalable solutions to the computational, communication and load balancing challenges.

\section*{Acknowledgments}
We gratefully acknowledge generous support from the Intel Parallel Computing Center (IPCC) grant, NSF OCI-114932 (Career: Programming Environments and Runtime for Data Enabled Science), CIF-DIBBS 143054: Middleware and High Performance Analytics Libraries for Scalable Data Science, NSF EAGRER grant, 
NSF Bigdata grant and DTRA CNIMS grant.  We appreciate the support from IU PHI, FutureSystems team and ISE Modelling and Simulation Lab.  

\bibliographystyle{IEEEtran}
\bibliography{reference}

\begin{thebibliography}{10}
\providecommand{\url}[1]{#1}
\csname url@samestyle\endcsname
\providecommand{\newblock}{\relax}
\providecommand{\bibinfo}[2]{#2}
\providecommand{\BIBentrySTDinterwordspacing}{\spaceskip=0pt\relax}
\providecommand{\BIBentryALTinterwordstretchfactor}{4}
\providecommand{\BIBentryALTinterwordspacing}{\spaceskip=\fontdimen2\font plus
\BIBentryALTinterwordstretchfactor\fontdimen3\font minus
  \fontdimen4\font\relax}
\providecommand{\BIBforeignlanguage}[2]{{%
\expandafter\ifx\csname l@#1\endcsname\relax
\typeout{** WARNING: IEEEtran.bst: No hyphenation pattern has been}%
\typeout{** loaded for the language `#1'. Using the pattern for}%
\typeout{** the default language instead.}%
\else
\language=\csname l@#1\endcsname
\fi
#2}}
\providecommand{\BIBdecl}{\relax}
\BIBdecl

\bibitem{chen:icdm16}
X.~Chen and J.~C.~S. Lui, ``Mining graphlet counts in online social networks,''
  in \emph{{{ICDM}}}, 2016, pp. 71--80.

\bibitem{milo2002network}
R.~Milo, S.~Shen-Orr, S.~Itzkovitz, N.~Kashtan, D.~Chklovskii, and U.~Alon,
  ``Network motifs: simple building blocks of complex networks,''
  \emph{Science}, vol. 298, no. 5594, p. 824, 2002.

\bibitem{khan:sigmod11}
A.~Khan, N.~Li, X.~Yan, Z.~Guan, S.~Chakraborty, and S.~Tao, ``Neighborhood
  based fast graph search in large networks,'' in \emph{{{SIGMOD}}}, New York,
  NY, USA, 2011, pp. 901--912.

\bibitem{bressan_counting_2017}
M.~Bressan, F.~Chierichetti, R.~Kumar, S.~Leucci, and A.~Panconesi, ``Counting
  {Graphlets}: {Space} vs {Time},'' in \emph{{{WSDM}}}, 2017, pp. 557--566.

\bibitem{vassilevska2009finding}
V.~Vassilevska and R.~Williams, ``Finding, minimizing, and counting weighted
  subgraphs,'' in \emph{{{STOC}}}, 2009, pp. 455--464.

\bibitem{flum2004parameterized}
J.~Flum and M.~Grohe, ``The parameterized complexity of counting problems,''
  \emph{SIAM Journal on Computing}, vol.~33, no.~4, pp. 892--922, 2004.

\bibitem{curticapean2014complexity}
R.~Curticapean and D.~Marx, ``Complexity of counting subgraphs: Only the
  boundedness of the vertex-cover number counts,'' in \emph{{{FOCS}}}.\hskip
  1em plus 0.5em minus 0.4em\relax IEEE, 2014, pp. 130--139.

\bibitem{alon_color-coding_1995}
N.~Alon, R.~Yuster, and U.~Zwick, ``Color-coding,'' \emph{J. ACM}, vol.~42,
  no.~4, pp. 844--856, Jul. 1995.

\bibitem{koutis:icalp08}
I.~Koutis, ``Faster algebraic algorithms for path and packing problems,'' in
  \emph{Proc. ICALP}, 2008.

\bibitem{williams2009finding}
R.~Williams, ``Finding paths of length k in o∗(k2) time,'' \emph{Information
  Processing Letters}, vol. 109, no.~6, pp. 315--318, 2009.

\bibitem{bjorklund:esa14}
A.~Bjorklund, P.~Kaski, and L.~Kowalik, ``Fast witness extraction using a
  decision oracle,'' in \emph{Proc. ESA}, 2014.

\bibitem{zhao_sahad:_2012}
Z.~Zhao, G.~Wang, A.~R. Butt, M.~Khan, V.~A. Kumar, and M.~V. Marathe, ``Sahad:
  Subgraph analysis in massive networks using hadoop,'' in \emph{{{IPDPS}}},
  2012, pp. 390--401.

\bibitem{slota_parallel_2015}
G.~M. Slota and K.~Madduri, ``Parallel color-coding,'' \emph{Parallel
  Computing}, vol.~47, pp. 51--69, 2015.

\bibitem{ekanayake:ipdps18}
S.~Ekanayake, J.~Cadena, U.~Wickramasinghe, and A.~Vullikanti, ``Midas:
  Multilinear detection at scale,'' in \emph{Proc. IPDPS}, 2018.

\bibitem{chakaravarthy_subgraph_2016}
V.~T. Chakaravarthy, M.~Kapralov, P.~Murali, F.~Petrini, X.~Que, Y.~Sabharwal,
  and B.~Schieber, ``Subgraph {{Counting}}: {{Color Coding Beyond Trees}},'' in
  \emph{{{IPDPS}}}, May 2016, pp. 2--11.

\bibitem{chen_benchmarking_2017}
L.~Chen, B.~Peng, B.~Zhang, T.~Liu, Y.~Zou, L.~Jiang, R.~Henschel, C.~Stewart,
  Z.~Zhang, E.~Mccallum, T.~Zahniser, O.~Jon, and J.~Qiu, ``Benchmarking
  {{Harp}}-{{DAAL}}: {{High Performance Hadoop}} on {{KNL Clusters}},'' in
  \emph{{{IEEE Cloud}}}, Honolulu, Hawaii, US, Jun. 2017.

\bibitem{harpdaalweb}
{Indiana University}, ``{Harp-DAAL official website},''
  \url{https://dsc-spidal.github.io/harp}, 2018, online; Accessed: 2018-01-21.

\bibitem{Zhang2015a}
B.~Zhang, Y.~Ruan, and J.~Qiu, ``Harp: {{Collective}} communication on
  {{Hadoop}},'' in \emph{{{IC2E}}}, 2015, pp. 228--233.

\bibitem{Zhang2016}
B.~Zhang, B.~Peng, and J.~Qiu, ``High performance {{LDA}} through collective
  model communication optimization,'' \emph{Procedia Computer Science},
  vol.~80, pp. 86--97, 2016.

\bibitem{peng_harplda+:_2017}
B.~Peng, B.~Zhang, L.~Chen, M.~Avram, R.~Henschel, C.~Stewart, S.~Zhu,
  E.~Mccallum, L.~Smith, T.~Zahniser, J.~Omer, and J.~Qiu, ``{HarpLDA}+:
  {Optimizing} latent dirichlet allocation for parallel efficiency,'' in
  \emph{2017 {IEEE} {International} {Conference} on {Big} {Data} ({Big}
  {Data})}, Dec. 2017, pp. 243--252.

\bibitem{inteldaalweb}
{Intel Corporation}, ``{The Intel Data Analytics Acceleration Library (Intel
  DAAL)},'' \url{https://github.com/intel/daal}, 2018, online; accessed
  2018-01-21.

\bibitem{hockney_communication_1994}
R.~W. Hockney, ``The communication challenge for {MPP}: Intel paragon and meiko
  {CS}-2,'' vol.~20, no.~3, pp. 389--398.

\bibitem{barrett_generation_2009}
C.~L. Barrett, R.~J. Beckman, M.~Khan, V.~S.~A. Kumar, M.~V. Marathe, P.~E.
  Stretz, T.~Dutta, and B.~Lewis, ``Generation and analysis of large synthetic
  social contact networks,'' in \emph{{{WSC}}}, Dec. 2009, pp. 1003--1014.

\bibitem{snapnets}
J.~Leskovec and A.~Krevl, ``{SNAP Datasets}: {Stanford} large network dataset
  collection,'' \url{http://snap.stanford.edu/data}, Jun. 2014.

\bibitem{yang_defining_2012}
J.~Yang and J.~Leskovec, ``Defining and {{Evaluating Network Communities
  Based}} on {{Ground}}-{{Truth}},'' in \emph{{{ICDM}}}, Dec. 2012, pp.
  745--754.

\bibitem{cha_measuring_2010}
M.~Cha, H.~Haddadi, F.~Benevenuto, and K.~P. Gummadi, ``Measuring {{User
  Influence}} in {{Twitter}}: {{The Million Follower Fallacy}}.'' in
  \emph{{{ICWSM}}}, vol.~14, Jun. 2010.

\bibitem{davis_university_2011}
T.~A. Davis and Y.~Hu, ``The {{University}} of {{Florida Sparse Matrix
  Collection}},'' \emph{ACM Trans. Math. Softw.}, vol.~38, no.~1, pp.
  1:1--1:25, Dec. 2011.

\bibitem{chakrabarti_r-mat:_2004}
D.~Chakrabarti, Y.~Zhan, and C.~Faloutsos, ``R-{{MAT}}: {{A}} recursive model
  for graph mining,'' in \emph{{{SIAM}}}, vol.~6, Apr. 2004.

\bibitem{kowaluk:soda11}
M.~Kowaluk, A.~Lingas, and E.-M. Lundell, ``Counting and detecting small
  subgraphs via equations and matrix multiplication,'' in \emph{{{SODA}}},
  2011, pp. 1468--1476.

\bibitem{zhao_subgraph_2010}
Z.~Zhao, M.~Khan, V.~A. Kumar, and M.~V. Marathe, ``Subgraph enumeration in
  large social contact networks using parallel color coding and streaming,'' in
  \emph{{{ICPP}}}, 2010, pp. 594--603.

\bibitem{slota_fast_2013}
G.~M. Slota and K.~Madduri, ``Fast approximate subgraph counting and
  enumeration,'' in \emph{{{ICPP}}}, 2013, pp. 210--219.

\bibitem{slota_complex_2014}
------, ``Complex network analysis using parallel approximate motif counting,''
  in \emph{{{IPDPS}}}, 2014, pp. 405--414.

\bibitem{alon:talg10}
N.~Alon and S.~Gutner, ``Balanced families of perfect hash functions and their
  applications,'' \emph{ACM Trans. Algorithms}, vol.~6, no.~3, pp. 54:1--54:12,
  Jul. 2010.

\end{thebibliography}

\end{document}